\documentclass[aps,pra,preprint,groupedaddress,showpacs]{revtex4-1}
\usepackage{subfigure} 
\usepackage{graphicx} 
\usepackage{dcolumn}  
\usepackage{bm}
\usepackage{amsfonts}
\usepackage{makeidx}
\begin{document}
\newcommand{\greeksym}[1]{{\usefont{U}{psy}{m}{n}#1}}
\newcommand{\rmssmu}{\mbox{\scriptsize{\greeksym{m}}}}
\newcommand{\rmsstau}{\mbox{\scriptsize{\greeksym{t}}}}
\newcommand{\rmssgamma}{\mbox{\scriptsize{\greeksym{g}}}}
\newcommand{\rmmu}{\mbox{\greeksym{m}}}
\newcommand{\rmtau}{\mbox{\greeksym{t}}}
\newcommand{\rmalpha}{\mbox{\greeksym{a}}}
\newcommand{\rmssalpha}{\mbox{\scriptsize{\greeksym{a}}}}
\newcommand{\rmpi}{\mbox{\greeksym{p}}}
\newcommand{\rmsspi}{\mbox{\scriptsize{\greeksym{p}}}}
\newcommand{\rmphi}{\mbox{\greeksym{f}}}
\def\lbar{\lambda\hskip-4.5pt\vrule height4.6pt depth-4.3pt width4pt}
\def\hbar{h\hskip-5pt\vrule height6.6pt depth-6.3pt width4pt}
\def\cU{{\cal U}}
\def\cV{{\cal V}}
\def\cW{{\cal W}}
\def\iD{{\it \Delta}}
\def\iG{{\it \Gamma}}
\def\iL{{\it \Lambda}}
\def\iX{{\it \Xi}}
\def\rC{{\rm C}}
\def\rI{{\rm I}}
\def\rL{{\rm L}}
\def\rXL{{\rm XL}}
\def\rN{{\rm N}}
\def\rP{{\rm P}}
\def\rS{{\rm S}}
\def\rT{{\rm T}}
\def\rX{{\rm X}}
\def\rp{{\rm p}}
\def\rq{{\rm q}}
\def\rr{{\rm r}}
\def\ru{{\rm u}}
\def\rd{{\rm d}}
\def\re{{\rm e}}
\def\rf{{\rm f}}
\def\rg{{\rm g}}
\def\ri{{\rm i}}
\def\0{{\bm 0}}
\def\m{{\mbox{-}}}
\def\p{{+\!}}
\newcommand{\as}{{^\ast\!}}
\def\fr#1#2{{\textstyle{\frac{#1}{#2}}}}
\def\fl#1#2{{\frac{\textstyle #1}{\textstyle #2}}}
\def\inv#1{}  
\preprint{Version 2.0}

\title{Bound-state field theory approach to proton structure effects  in muonic hydrogen}

\author{Peter J. Mohr}
\email[]{mohr@nist.gov}
\affiliation{National Institute of Standards and Technology, Gaithersburg, MD 20899-8420}

\author{J. Griffith}
\email[]{jgriff8@nd.edu}
\affiliation{Department of Physics, University of Notre Dame, Notre Dame, IN 46556}

\author{J. Sapirstein}
\email[]{jsapirst@nd.edu}
\affiliation{Department of Physics,
University of Notre Dame, Notre Dame, IN 46556}

\begin{abstract}
A bound-state field theory approach to muonic hydrogen is set up using a
variant of the Furry representation in which the lowest-order
Hamiltonian describes a muon in the presence of a point Coulomb field,
but the origin of the binding field is taken to be three charged quarks
in the proton which are modeled as Dirac particles that move freely
within a spherical well. Bound-state field theory techniques are used to
evaluate one- and two-photon effects. Particular attention is paid to
two-photon exchange diagrams, which include the effect of proton
polarizability. In addition the modification of the electromagnetic self
energy of the proton by the electric field of the muon is examined.
Finally, the model is used to carry out a calculation of the static
electric polarizability of the proton. 
\end{abstract}


\maketitle

\section{introduction}

One of the simplest ways to model the proton is as three very light
quarks confined in a spherical well. Choosing the radius of the well to
be $1.2$ fm leads to moderately good agreement with experiment for its
electromagnetic properties, such as the charge radius, magnetic moment,
and static electric and magnetic polarizabilities. This model, a
simplified version of the MIT bag model \cite{MIT}, will be referred to
in the following as the static-well model.  It allows an alternative
approach to the calculation of the electromagnetic properties of the
proton, generally treated with methods quite different in character,
that uses the methods of conventional bound-state QED. The latter theory
is characterized by wave functions that satisfy the Dirac equation in an
external field along with electron propagators defined in terms of the
same field. When the external field is that of a point Coulomb source, a
modification of the interaction representation introduced by Furry
\cite{Furry} allows a systematic Feynman diagram treatment of radiative
corrections. This approach can also be applied to many-electron systems,
and a Feynman diagram treatment of electron-electron interactions is
also possible. As will be explained below, the present paper is
patterned on a calculation of these interactions in heliumlike ions
involving two-photon exchange \cite{heliumexcited}.

The approach we will use in this paper was applied some time ago
\cite{oldprl} to the computation of the electromagnetic self energy of
the proton and neutron. In that work, both the effect of exchange of a
photon between quarks along with the electromagnetic self energy of the
quarks were evaluated and found to sum to $0.53$ MeV for the proton and
$-0.28$ MeV for the neutron for the case of nearly zero-mass quarks. The
fact that the proton is lighter than the neutron remains explained by
the fact that the down quark is heavier than the up quark, but it is of
note that the electromagnetic correction to the mass splitting, $-0.8$
MeV, is the same order-of-magnitude as the neutron-proton mass
difference, $1.2$ MeV.

The proton can be studied with electron-scattering experiments, which
have a long history of providing information about its properties, in
particular the root-mean-square (rms) radius, $r_{\rm p}$. The proton
size has recently received considerable attention because of unexpected
results for the $2s_{1/2}-2p_{3/2}$ transition energy of muonic hydrogen
\cite{Pohl}. The issue of determining $r_{\rm p}$ from scattering data
can be problematic, as extrapolating the slope of the Dirac form factor
to $q^2=0$ involves a number of assumptions \cite{Hill-Paz}. An
alternative approach is to determine the proton size by doing precise
measurements of atomic transitions that are sensitive to the effect of
the size. The 2010 CODATA result \cite{CODATA} in fact uses this
procedure with hydrogen and deuterium, where the experiment and theory
are so accurate that the proton size can be inferred with an accuracy
comparable to that available from scattering experiments as of 2010.

Because of its smaller size, muonic hydrogen has long been recognized as
a system whose spectrum could be used to determine a much more accurate
rms radius of the proton than that obtained from hydrogen and deuterium,
but the associated experimental obstacles have only recently been
overcome. While indeed much more accurate, the result of Ref.
\cite{Pohl} for the proton size,
\begin{equation}
r_{\rm p} = 0.841\,84(67)\, {\rm fm} ,
\end{equation}
is significantly smaller than the CODATA result,
\begin{equation}
r_{\rm p} = 0.8768(69)\, {\rm fm}.
\end{equation}
This discrepancy is referred to as the muonic hydrogen puzzle.

One possible explanation of the puzzle involves the electromagnetic
structure of the proton, and the largest theoretical uncertainty comes
from an effect called proton polarizability. This is generally evaluated
by relating the energy shift to forward virtual photon-proton
scattering. The amplitude describing this scattering, $T^{\mu \nu}(\nu,
q^2)$,  can then be related to proton form factors through dispersion
relations.  A recent paper that covers all contributions to muonic
hydrogen with particular attention to proton polarizability is
Ref.~\cite{Pohltheory}; in the conclusion, we compare our results to
results quoted in that paper.  A number of issues involving convergence
of the dispersion theory integrals and the need for experimental data
complicate that approach.  The purpose of the present paper is to
provide an alternative analysis patterned after bound-state field theory
calculations in atomic physics.  This will be done by using the
static-well model of the proton together with standard bound-state QED.
As we will show, there is a natural way of setting up a consistent QED
calculation for hydrogen and muonic hydrogen, with the proton treated as
a bound state of three quarks interacting with an electron or a muon,
that requires no scattering information for its predictions; rather it
depends only on the radius of the well.

Regarding the proton as three relativistic particles confined to a small
volume is closely analogous to treating three electrons in
highly-charged ions, where the electrons for large nuclear charge $Z$
are quite relativistic and the ion has a size of $1/Z$ Bohr radius. This
problem has recently been addressed with techniques similar to those used
for heliumlike ions mentioned above \cite{heliumexcited}, and have been
shown to provide an accurate description of these ions
\cite{stpeteli,llnllithium}, the spectra of which have been measured
with high accuracy \cite{llnlexp}. 

In these calculations almost all of the important physics is described
by Feynman diagrams with one or two photons. The same turns out to hold
for the present calculation, though in this paper, while we will show
all relevant diagrams, we concentrate our attention on two effects
dependent on proton structure, the polarizability of the proton and the
screening of the proton electromagnetic self energy. 

Our model of the proton is extremely simple, but there are three reasons
we have chosen it.  The first is that proton structure effects are
generally very small, with even the largest, the effect of its finite
size, accounting for about 2 percent of the transition energy in muonic
hydrogen. Thus even a crude determination of a proton structure effect
will have a small relative theoretical error. The second is that while
the proton polarizability correction has been evaluated with other
methods, a contribution we term the proton Lamb shift has not, and the
results presented here may stimulate more sophisticated calculations.
The final reason is that mentioned above, to explore a method of
calculating the effect of proton structure on atomic energy levels that
does not require the use of dispersion theory. 

We will in the following consider the effect of proton structure on both
electronic and muonic hydrogen. Because our formalism does not include
recoil, we will present results in terms of the electron mass $m_\re$
and the muon mass $m_{\rmssmu}$ even though reduced-mass effects on the
latter are about 10 percent. When we give a general formula, we refer to
a lepton with mass $m_l$. The state of the lepton, in practice either
$2s$  or $2p_{3/2}$, will be denoted $v$, and the index $m$ will be used
for sums over intermediate leptonic states: for the corresponding case
of quarks, we use $\rg$ to denote a ground-state quark and reserve $n$
for sums over intermediate quark states.

The plan of the paper is as follows. We begin in section II with a
quantum mechanical (QM) treatment of the shift in energy levels arising
from the perturbation of replacing the potential of a point proton
Coulomb field with that of a general distribution of charge $\rho(\bm
r)$.  This perturbation theory is evaluated through second order. In
section III we turn to a quantum-field-theoretic approach to the problem
in the context of the static-well model. We do this by modifying the
standard Furry representation \cite{Furry} through forcing the
lowest-order Hamiltonian to be same as that used in that representation,
but having the quarks in the proton provide the Coulomb field instead of
assuming a point source.  This requires the introduction of a new term
in the interaction Hamiltonian we call the counter term, the effects of
which however are quite simple to evaluate. We also define the
static-well model and briefly review the calculation of the proton
electromagnetic self energy. In section IV we use bound-state field
theory to treat one photon exchange, and show that the results agree
with the first order QM energy. We note here that in this paper we use
the Coulomb gauge and treat only Coulomb photons for all exchanged
photons. For the QM approach this corresponds to ignoring magnetic
effects, and in field theory to leaving out transverse-photon exchange.
In section V we then turn to two-photon exchange diagrams, which we
break into two classes, one in which only one photon attaches to the
lepton, with the other being emitted and reabsorbed in the proton,  and
a second in which each photon is exchanged between the lepton and a
quark.  In Section Va we treat the first class, which has no QM analog,
and present a calculation of the contribution, which we call the proton
Lamb shift. In Section Vb we treat the second class, but again make a
breakup of the diagrams into firstly a part in which the proton is left
unchanged, and secondly a part where it is excited. (In our model this
means that in the spectral representation of the quark propagator it is
either saturated with the $1s$ state, or else that state is excluded).
The first part will be shown to correspond exactly to the second-order
QM energy. The second part, the proton polarizability, is then
evaluated. The related calculation of the proton's static electric
polarizability is carried out in Section VI, and it is shown that in the
$\kappa=1$ angular momentum channel a complete cancellation between
positive- and negative-energy state terms occurs, leaving only
contributions from the $\kappa=-2$ intermediate states. In the
conclusion we compare our results to the results of other calculations
and describe directions for future progress. 

\section{Perturbation theory of finite-nuclear-size effects}

We consider a central potential for hydrogen or muonic hydrogen coming
from a finite charge distribution $\rho(\bm x)$, normalized to unity.
The corresponding static potential is
\begin{eqnarray} 
V(\bm x) &=& -Z\alpha\int\rd \bm x^\prime \,
\frac{\rho(\bm x^\prime)}{|\bm x - \bm x^\prime|}.
\end{eqnarray}
While we have $Z=1$, the following discussion can also be applied to the
case $Z \neq 1$. We start with a point-Coulomb binding field, so this
distribution leads to the perturbation
\begin{eqnarray}
\delta V(\bm x) &=& 
 -Z\alpha\int\rd \bm x^\prime \,
\frac{\rho(\bm x^\prime)-\delta(\bm x^\prime)}{|\bm x - \bm x^\prime|}.
\label{Vparts}
\end{eqnarray}

The first-order correction,
\begin{eqnarray}
E^{(1)} &=&  \int d \bm x
\phi_v^\dagger(\bm x)\,\delta V(\bm x)
\,\phi_v(\bm x),
\label{firstorderpt}
\end{eqnarray}
is valid for either a relativistic or nonrelativistic calculation.  We
first consider the nonrelativistic limit. Then in leading order the wave
functions may be replaced by their value at the origin and the
first-order energy is
\begin{eqnarray}
E^{(1)}_0 &=& 
 - Z\alpha \,
|\phi_v(0)|^2 \int\rd\bm x 
\int\rd \bm x^\prime \,\frac{
\rho(\bm x^\prime) - \delta(\bm x^\prime)}{|\bm x - \bm x^\prime|}
\nonumber\\[10 pt]&=& - Z\alpha \,
|\phi_v(0)|^2 \int\rd \bm x^\prime \,
 4\pi\left(\frac{1}{u^2} - \frac{1}{6}\,\bm x^{\prime 2} + \dots\right)
\left[\rho(\bm x^\prime) - \delta(\bm x^\prime)\right]
\nonumber\\[10 pt]&=&  \frac{2\pi Z\alpha}{3} \,
|\phi_v(0)|^2 \int\rd \bm x \,
\bm x^{2} \rho(\bm x)
= \frac{2(Z\alpha)^4}{3n^3}\,m_l^3
\int\rd \bm x \,
\bm x^{2} \rho(\bm x),
\label{rms1}
\end{eqnarray}
where in the last step we have assumed $v$ to be an $ns$-state.

The integral over $d \bm x$ has been carried out using
a cutoff procedure which we now describe. We introduce a parameter $u$,
understood to ultimately be taken to zero, and work with the basic identity  
\begin{equation}
\int\rd \bm x \, \frac{1}{|\bm x_2 - \bm x|} \,
\frac{\re^{-u|\bm x - \bm x_1|}}{|\bm x - \bm x_1|} = {4 \pi \over u^2} {1-e^{-u|\bm x_2 - \bm x_1|} \over |\bm x_2 - \bm x_1|},
\end{equation}
which for small $u$ has the expansion
\begin{eqnarray}
\int\rd \bm x \, \frac{1}{|\bm x_2 - \bm x|} \,
\frac{\re^{-u|\bm x - \bm x_1|}}{|\bm x - \bm x_1|}
&=& 4\pi\left(\frac{1}{u} - \frac{1}{2}\,|\bm x_2 - \bm x_1| + \frac{u}{6}\,|\bm x_2 - \bm x_1|^2
-\frac{u^2}{24}\,|\bm x_2 - \bm x_1|^3 +
\dots\right) 
\label{eq:int1}
\end{eqnarray}
By differentiating once or twice with respect to $u$, we have
\begin{equation}
\int\rd \bm x \, 
\frac{\re^{-u|\bm x - \bm x_1|}}{|\bm x_2 - \bm x|}
= 4\pi\left(\frac{1}{u^2} - \frac{1}{6}\,|\bm x_2 - \bm x_1|^2 + {\cal O}(u) \right)
\label{eq:rint1}
\end{equation}
and
\begin{equation}
\int\rd \bm x \, 
\frac{\re^{-u|\bm x - \bm x_1|}}{|\bm x_2 - \bm x|}
\, |\bm x - \bm x_1|
= 4\pi\left(\frac{2}{u^3} - \frac{1}{12}\,|\bm x_2 - \bm x_1|^3 +
{\cal O}(u)\right).
\label{eq:rint2}
\end{equation}

We have used Eq.~(\ref{eq:rint1}) in the derivation of Eq.~(\ref{rms1}),
and use Eq.~(\ref{eq:rint2}) to evaluate the correction coming from the
variation of the wave function to leading order. For S states, this
arises from
\begin{eqnarray}
|\phi_v(\bm x)|^2 &=& |\phi_v(0)|^2 \left(1 - 2Z\alpha m_l|\bm x| +
\dots\right),
\end{eqnarray}
which yields an additional contribution of
\begin{eqnarray}
E^{(1)}_1 
& = & 
2(Z\alpha)^2 m_l\,
|\phi_v(0)|^2 \int\rd\bm x \, |\bm x|
\int\rd \bm x^\prime \,\frac{
\rho(\bm x^\prime) - \delta(\bm x^\prime)}{|\bm x - \bm x^\prime|}
\nonumber\\[10 pt]&=& 2 (Z\alpha)^2 m_l\,
|\phi_v(0)|^2 \int\rd \bm x^\prime \,
 4\pi\left(\frac{2}{u^3} - \frac{1}{12}\,|\bm x^\prime|^3 + \dots\right)
\left[\rho(\bm x^\prime) - \delta(\bm x^\prime)\right]
\nonumber\\[10 pt]&=& - \frac{2\pi (Z\alpha)^2}{3} \, m_l \,
|\phi_v(0)|^2 \int\rd \bm x \,
|\bm  x|^3 \rho(\bm x)
= -\frac{2(Z\alpha)^5}{3n^3}\,m_l^4
\int\rd \bm x \,
\bm x^{3} \rho(\bm x).
\label{rms2}
\end{eqnarray}
This term cancels a corresponding term in second-order perturbation
theory, which is given by the standard form
\begin{eqnarray}
E^{(2)} &=& \int\rd\bm x_2 \int\rd\bm x_1\,
\phi_v^\dagger(\bm x_2)\,\delta V(\bm x_2)
\sum_{m\ne v}\frac{\phi_m(\bm x_2)\,\phi_m^\dagger(\bm x_1)}
{\epsilon_v - \epsilon_m}\,\delta V(\bm x_1)\,\phi_v(\bm x_1).
\label{secondorderpt}
\end{eqnarray}
The sum over terms involving $m$ is $-1$ times the reduced Green
function.  This expression is again valid for either a relativistic or
nonrelativistic calculation.  The correction to the potential is only
non-zero outside the nucleus, which means that the wave functions and
reduced Green function are evaluated for small arguments, because the
Bohr radius for both the muon and electron is large compared to the
nuclear size.  We again take the nonrelativistic limit. Then the wave
functions may be evaluated at the origin and the reduced Green function
may be replaced by the nonrelativistic free Green function to give
\begin{eqnarray}
E^{(2)}_0 &=& - \frac{m_l}{2\pi}\,|\phi_v(0)|^2
\int\rd\bm x_2 \int\rd\bm x_1\,
\delta V(\bm x_2) \,
\frac{1}{|\bm x_2 - \bm x_1|} \,
\,\delta V(\bm x_1)
\nonumber\\[10 pt]
&=& - \frac{(Z\alpha)^2m_l}{2\pi}\,|\phi_v(0)|^2
\int\rd\bm x_2^\prime \int\rd\bm x_1^\prime \,
\nonumber\\[10 pt]&&\times
\int\rd\bm x_2 \int\rd\bm x_1\,
\frac{\rho(\bm x_2^\prime)-\delta(\bm x_2^\prime)}
{|\bm x_2 - \bm x_2^\prime|} \,
\frac{1}{|\bm x_2 - \bm x_1|} \,
\frac{\rho(\bm x_1^\prime)-\delta(\bm x_1^\prime)}
{|\bm x_1 - \bm x_1^\prime|} \,.
\end{eqnarray}
Introducing cutoffs allows us to carry out the integrals over the
unprimed variables with the formulas given above, resulting in
\begin{eqnarray}
E^{(2)}_0 &=& 
(Z\alpha)^2m_l\,|\phi_v(0)|^2
\int\rd\bm x_2^\prime \int\rd\bm x_1^\prime \,
\nonumber\\[10 pt]&&\times
 \int\rd\bm x_1\,
\left[\rho(\bm x_2^\prime)-\delta(\bm x_2^\prime)\right]
|\bm x_2^\prime - \bm x_1| \,
\frac{\rho(\bm x_1^\prime)-\delta(\bm x_1^\prime)}
{|\bm x_1 - \bm x_1^\prime|} \,,
\end{eqnarray}
which yields 
\begin{eqnarray}
E^{(2)}_0 &=&  - \frac{\pi(Z\alpha)^2m_l\,|\phi_v(0)|^2}{3}
\Bigg[
\int\rd\bm x_2 \int\rd\bm x_1 \,
\rho(\bm x_2)
|\bm x_2 - \bm x_1|^3 
\rho(\bm x_1) 
 -2 \int\rd\bm x \,
|\bm x|^3 \,
\rho(\bm x)
\Bigg]
\qquad \nonumber \\
 &=&  - \frac{(Z\alpha)^5}{3 n^3}\,m_l^4
\Bigg[
\int\rd\bm x_2 \int\rd\bm x_1 \,
\rho(\bm x_2)
|\bm x_2 - \bm x_1|^3 
\rho(\bm x_1) 
 -2 \int\rd\bm x \,
|\bm x|^3 \,
\rho(\bm x)
\Bigg].
\qquad\label{eq:sopt}
\end{eqnarray}
As alluded to above, the second term in the square brackets in
Eq.~(\ref{eq:sopt}) is cancelled by Eq.~(\ref{rms2}). The first term in
the square brackets is the third Zemach moment, which we denote $\langle
r^3 \rangle_Z$. An interesting feature about this term is that it too is
cancelled by a term that arises when the nucleus is allowed to undergo
low-energy excitations (proton polarizability), though we will not use
this fact directly, and instead just evaluate the entire effect.  We
turn now to a field-theoretic approach based on the static-well model,
and begin by introducing a formalism for bound-state field theory.

\section{Formalism}

While the formalism we use here is to our knowledge novel, it is a
simple extension of the Furry representation  \cite{Furry}, which we now
briefly review. We will use this representation both for leptons and
quarks, and begin by describing how it is used for the former. The full
QED Hamiltonian used for describing the scattering of free leptons is
$H=H_0 + H_\rI$, with ($\{\hbar\}=1,\{c\}=1,\{e\}=1$),
\begin{equation}
H_0 = \int \rd \bm x \, {\psi}^{\dagger}(x)
\left[ \bm\alpha \cdot \bm p + \beta m_l \right]  \psi(x)
\end{equation}
and
\begin{equation}
H_\rI = q_\re \int \rd \bm x \, \overline{\psi}(x) \gamma_{\mu} 
\psi(x) A^{\mu}(x),
\end{equation}
with $q_\re = -e$. We suppress normal ordering and the self-mass
counter term for simplicity. The Furry representation is used when $H_0$
is replaced by
\begin{equation}
\tilde{H}_0 = \int \rd \bm x \, {\psi}^{\dagger}(x)
\left[ \bm \alpha \cdot \bm p + \beta m_l - 
{Z \alpha \over |\bm x|}\right] \psi(x).
\end{equation}
This builds in a classical Coulomb field from an infinite mass proton.
Carrying out a unitary transformation to eliminate $\tilde{H}_0$ rather
than $H_0$ leads to the Furry representation in place of the interaction
representation. While the interaction Hamiltonian $H_\rI$ of a lepton
with photons keeps the same form, the lowest order spectrum now consists
of hydrogenic bound and scattering states, and the lepton Green function 
obeys the relation
\begin{equation}
\left(-\ri \bm \alpha \cdot \bm \nabla_{\!\bm x} + \beta m_l - 
{Z \alpha \over |\bm x|} -z\right) G(\bm x, \bm y; z) 
= \delta(\bm x - \bm y),
\end{equation}
which has the spectral representation
\begin{equation}
G(\bm x, \bm y; z) = \sum_m \frac{ \phi_m(\bm x) \phi^{\dagger}_m(\bm
y)}{\epsilon_m - z}.
\label{spectral}
\end{equation}
In the present case we treat proton structure using the static-well
model to specify the wave functions and Green functions of the
constituent quarks. However, the influence of the proton on a lepton in
a bound state cannot be treated perturbatively, so we need to build the
binding of the lepton into the formalism nonperturbatively. We do this
by modifying the breakup of the Hamiltonian given above to $H = (H_0 +
H_\rX) + (H_\rI - H_\rX)$. If we choose
\begin{equation}
H_\rX = -\int \rd \bm x \, \psi^{\dagger}(x) \,
\frac{Z \alpha}{|\bm x|}\, \psi(x) ,
\end{equation}
then this breakup is $H = \tilde{H}_0 + (H_\rI - H_\rX)$. In the
following we refer to $H_\rX$ as the counter term, though of course it is
to be distinguished from the electron mass counter term. We stress that
we do {\em not} assume that a classical Coulomb field is present.
However, because we have added a new term to the interaction
Hamiltonian, $\tilde{H}_0$ is unchanged from the usual Furry
representation, and the same wave functions and Green functions used in
Feynman diagram calculations in that representation can be used,
although extra Feynman diagrams involving $H_\rX$ need to be included. 

The use of the Furry representation must be extended to quarks in order
to account for the proton's Coulomb field. In this case $H_0$ and
$H_\rI$ are almost identical to the free case, but in $H_0$ we assume
the presence of a static well that confines the quarks, the details of
which are given below, so that the up and down quark fields are expanded
in terms of solutions to the Dirac equation in this well. The proton
then consists of the usual two up quarks and one down quark, and the sum
of their charges leads to the Coulomb field felt by the lepton.

Since the lowest-order problem describes the basic physics of the atom,
there has to be cancellation between diagrams in which a Coulomb photon
is exchanged between the lepton and the quarks in the proton and
diagrams with one counter term. In second order another cancellation
between two-photon exchange diagrams and diagrams involving one and two
counter terms must take place, and so on. Because the proton is now
modeled as a finite object, the cancellation will not be complete, and
we will identify the parts remaining after the near cancellation as the
subject of this paper, proton structure effects.  It is of  course vital
for this procedure to make sense that  the cancellation not only takes
place, but that the perturbation expansion converges.  We will present
results for the first and second terms in the expansion below. In
determining the order of the expansion we note that the counter term
$H_\rX$ is of the same order as two $H_\rI$'s.

The static-well model has well-known solutions which we show partly to
establish notation. We represent the solution to the free Dirac equation
in a spherically symmetric well, centered at the same origin as used in
Furry representation, by
\begin{equation}
\phi(\bm r) = \left( \begin{array}{c}  f_1(r) 
\chi_{\kappa \mu}(\bm{\hat{r}}) \\
i f_2(r) \chi_{-\kappa \mu}(\bm{\hat{r}}) \end{array} \right).
\end{equation}
Here $\chi_{\kappa \mu}(\bm{\hat{r}})$ is a spherical spinor and the
radial wavefunctions obey the equations
\begin{eqnarray}
\left({\partial \over \partial r} + {1 + \kappa \over r}\right) f_1(r) 
- \left[m(r)+E\right]f_2(r) & = & 0 \\
\left({\partial \over \partial r}  + {1 - \kappa \over r} \right)f_2(r) 
+ \left[E-m(r)\right]f_1(r) & = & 0.
\end{eqnarray}

Confinement is enforced by choosing $m(r)$ to be constant inside the
well and tending to infinity for $r>R$, which leads to the MIT bag model
boundary conditions \cite{MIT}.  The ground-state solution for the case
$m(r)=0$ for $r<R$, which will be used throughout this paper, is
\begin{eqnarray}
f_1(r) & = &  \frac{N}{r} \, \sin(wx) 
\nonumber \\
f_2(r) & = &  \frac{N}{r} \left[\cos(wx)-\frac{\sin(wx)}{wx}\right],
\end{eqnarray}
with $x=r/R$ and
\begin{equation}
N = {1 \over \sqrt{R}} \, \sqrt{{2{\it w}^2 \over 2{\it w}^2 + {\rm cos} 2{\it w} -1}}.
\end{equation}
Here $w=2.042\,787$ for the ground state, which gives $\epsilon_\rg =
335.9$ MeV for $R=1.2$ fm. When this wave function is used for the up
and down quarks the second and third moments, which can be calculated
analytically, are
\begin{equation}
\langle r^2 \rangle = R^2 \, \frac{ 2w^3-2w^2+4w-3}{6w^2(w-1)} 
                         = R^2\,0.531\,392
\end{equation}
and
\begin{equation}
\langle r^3 \rangle = R^3 \, \frac{2w^2-2w+3}{8w(w-1)} 
                          = R^3\,0.426\,041.
\end{equation} 
While an analytic form can be derived for the third Zemach moment, it is
lengthy and involves the sine integral, so we give only its numerical
value,
\begin{equation}
\langle r^3 \rangle_Z = R^3 \, 1.280\,621.
\end{equation}

The wave functions of the up and down quarks are identical in this zero
mass case, and we denote them as $\phi_\rg(\bm x)$.  This approximation
leads to the important simplification that the charge density of the
proton, even though it consists of three quarks, can be written in terms
of $\phi_\rg(\bm x)$,
\begin{equation}
\rho_\rg(\bm x) = \phi_\rg^{\dagger}(\bm x) \phi_\rg(\bm x).
\end{equation}
We will see that in our calculations of one- and two-Coulomb photon
exchange this density will enter in exactly the same way as it does in
Sec.~II in parts of the calculation, thereby reproducing the results of
that section with the static-well model charge density. Extra terms
arising from the field theory approach will be identified with
polarizability effects. 

Were we to use different masses, the charge densities of the up and down
quarks would differ, and in that case we would use
\begin{equation}
\rho_\rg(\bm x) = \frac{4}{3} \, \phi_\ru^{\dagger}(\bm x) \phi_\ru(\bm x)
- \frac{1}{3} \, \phi_\rd^{\dagger}(\bm x) \phi_\rd(\bm x).
\end{equation}
The $2s_{1/2}$ and $2p_{3/2}$ atomic wavefunctions for the lepton are
the standard Dirac-Coulomb solutions and will be denoted as $\phi_v(\bm
x)$. As proton structure effects are strongly suppressed for the
$2p_{3/2}$ case, even though we will continue to use $v$ in formulas, in
practice we will always assume $v=2s_{1/2}$.

The radius $R$ is the only variable in this calculation, and we will use
different values to study the $R$ dependence of what is by far the
largest proton structure correction, the finite size correction from
one-photon exchange. However, because all other proton structure effects
are much smaller, the value 1.2 fm is understood to be used for those
corrections.

For the calculation carried out here, which involves a quark propagating
in the well, we use the same kind of spectral decomposition as given in
Eq.~(\ref{spectral}),
\begin{equation}
G_\rq(\bm x, \bm y; z) = \sum_{n} \frac{ \phi_{n}(\bm x) 
{\phi_{n}}^{\dagger}(\bm y)}{\epsilon_n-z}.
\label{quarkprop}
\end{equation}
The sum over $m$ for the lepton and $n$ for the quark Green functions
can be carried out using the method of finite basis sets
\cite{bsplines}, which have been used extensively for atomic
calculations, with only minor modifications of the associated computer
code required for application to the quark Green function. This is
because the atomic calculations were set up in the same kind of
confining well as used here, but in that case only for the purpose of
discretizing the spectrum, with the well radius chosen to be much larger
than the atom or ion being considered. 

We will need the explicit form for the spin-up and spin-down proton wave
function, with the former being
\begin{eqnarray}
\left|\,p\,_{\uparrow} \right> & = & 
\frac{\epsilon_{ijk}}{\sqrt{72}}
\left[-2 b^{\dagger}_{id} b^{\dagger}_{ja} b^{\dagger}_{ka} 
+ b^{\dagger}_{ic} b^{\dagger}_{jb} b^{\dagger}_{ka}
+ b^{\dagger}_{ic} b^{\dagger}_{ja} b^{\dagger}_{kb}\right]\left|0 \right>.
\label{eq:qwf}
\end{eqnarray}
In Eq.~(\ref{eq:qwf}) $a$ and $b$ denote spin up and down states of an
up quark and $c$ and $d$ spin up and down states of a down quark;
$\epsilon_{ijk}$ is the Levi-Civita symbol, which makes the proton a
color singlet after the implicit sum over colors $ijk$ is carried out.
Because we are taking the up and down quark masses equal to zero, the
associated wave function $\phi_c(\bm x)$ can be replaced by
$\phi_a(\bm x)$ and $\phi_d(\bm x)$ by $\phi_b(\bm x)$, which
simplifies later formulas.  When the spin state of the quark is not
important, we simply use $\phi_\rg(\bm x)$.

Energy shifts are calculated with the use of $S$-matrix techniques, where 
we use Sucher's generalization of the Gell-Mann Low formula \cite{Sucher},
\begin{equation}
 \Delta E = \lim_{\epsilon \rightarrow 0, \, \lambda \rightarrow 1} 
 \frac{\ri \epsilon}{2}\, 
 \frac{\partial}{\partial \lambda} \ln 
 \left< v\left|\rT\!\left[ e^{-\ri \lambda H_\rI(\epsilon)}\right]\right|v
 \right>.
\end{equation}
Here $H_\rI(\epsilon)$ indicates that a factor $e^{-\epsilon |t|}$ is
included in the time integral over the Hamiltonian density in order to
adiabatically turn off the interaction at large positive and negative
times. The advantage of this formula is that the $S$-matrix can be
described with standard Feynman diagram techniques, with the adiabatic
factors usually trivially leading to a factor $1/\epsilon$ that cancels
the $\epsilon$ in the numerator of the above formula, though when we
deal with two-photon diagrams, the formalism is needed to cancel
disconnected diagrams. Details of how this works along with other
technical issues can be found in Ref.~\cite{heliumexcited}. That work
described a calculation of two-photon exchange diagrams contributing to
energy shifts of excited states of heliumlike ions, but the basic
approach is almost identical. The most important difference is that
while in that work $\left|v\right>$ describes two electrons, here it
describes one electron or muon and three quarks, and is given by
\begin{equation}
\left|v\right> = b_v^{\dagger} \left| p \right>.
\end{equation}

\begin{figure}[h]
  \begin{center}
    \subfigure[~Self-energy]
    {\label{Hlamb1}\includegraphics[width=7cm]{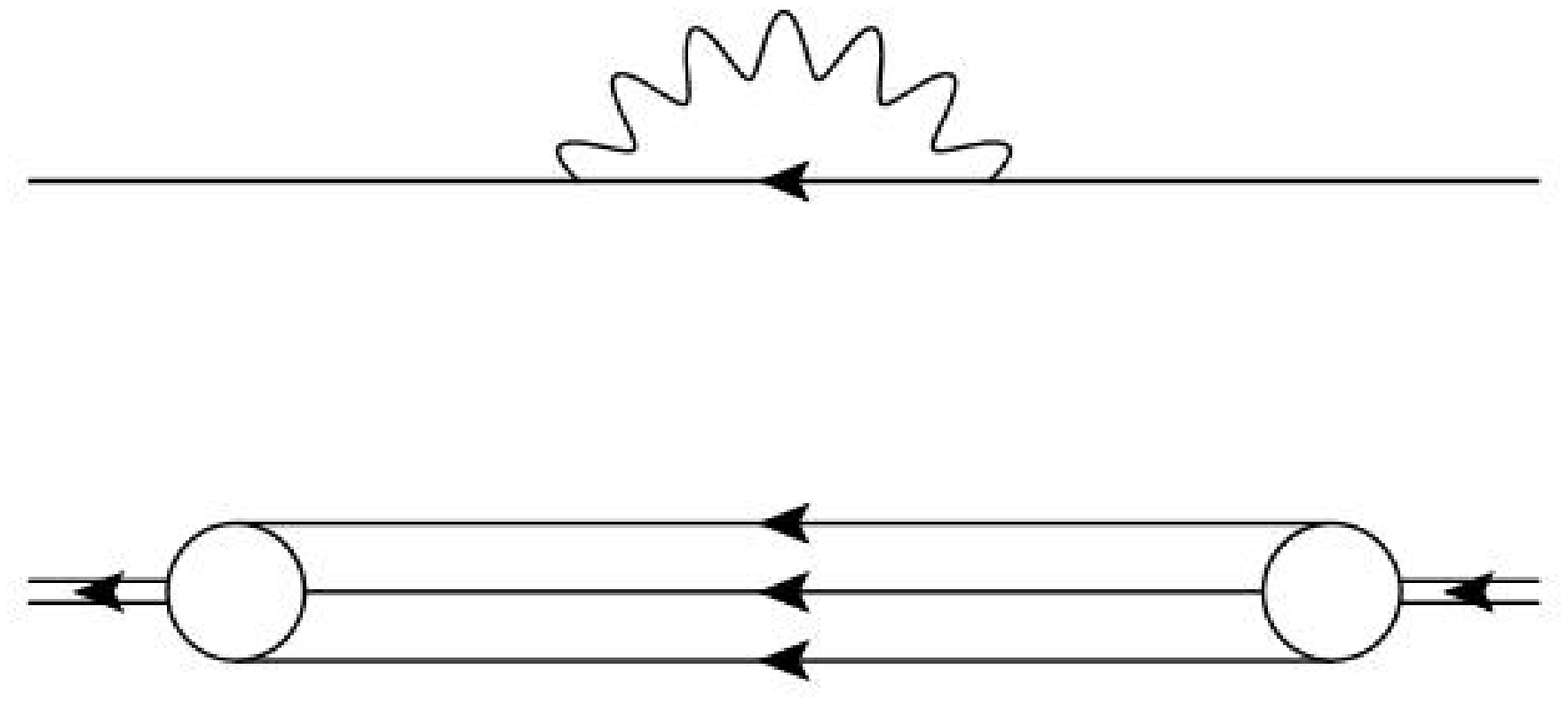}}
    \qquad
    \subfigure[~Vacuum polarization]
    {\label{Hlamb2}\includegraphics[width=7cm]{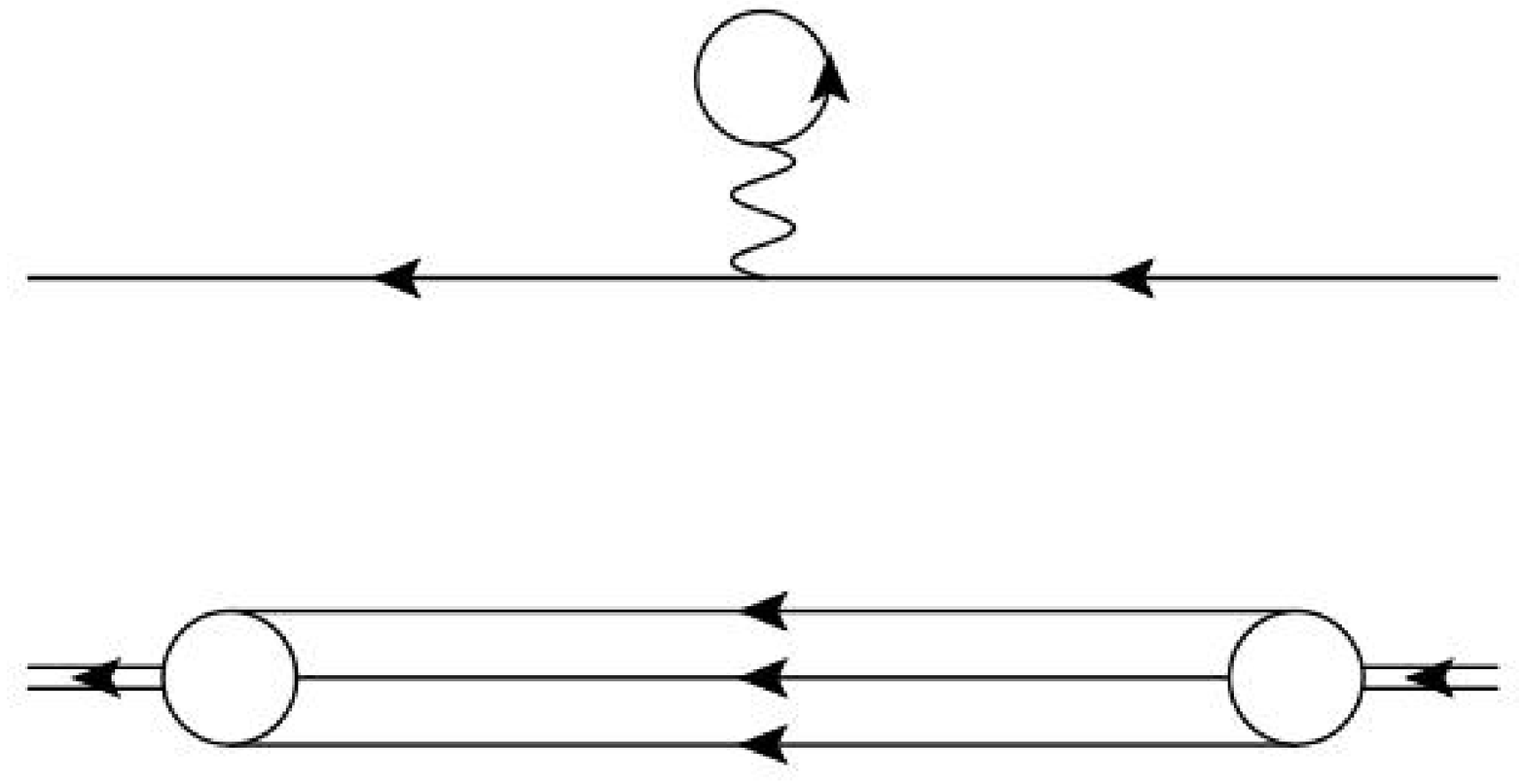}}
  \end{center}
  \caption{The two diagrams contributing to the Lamb shift in the hydrogenic 
    bound state.}
  \label{Hlamb}
\end{figure}

\begin{figure}[h]
  \begin{center}
    \subfigure[~Exchange]
    {\label{qlamb1}\includegraphics[width=7cm]{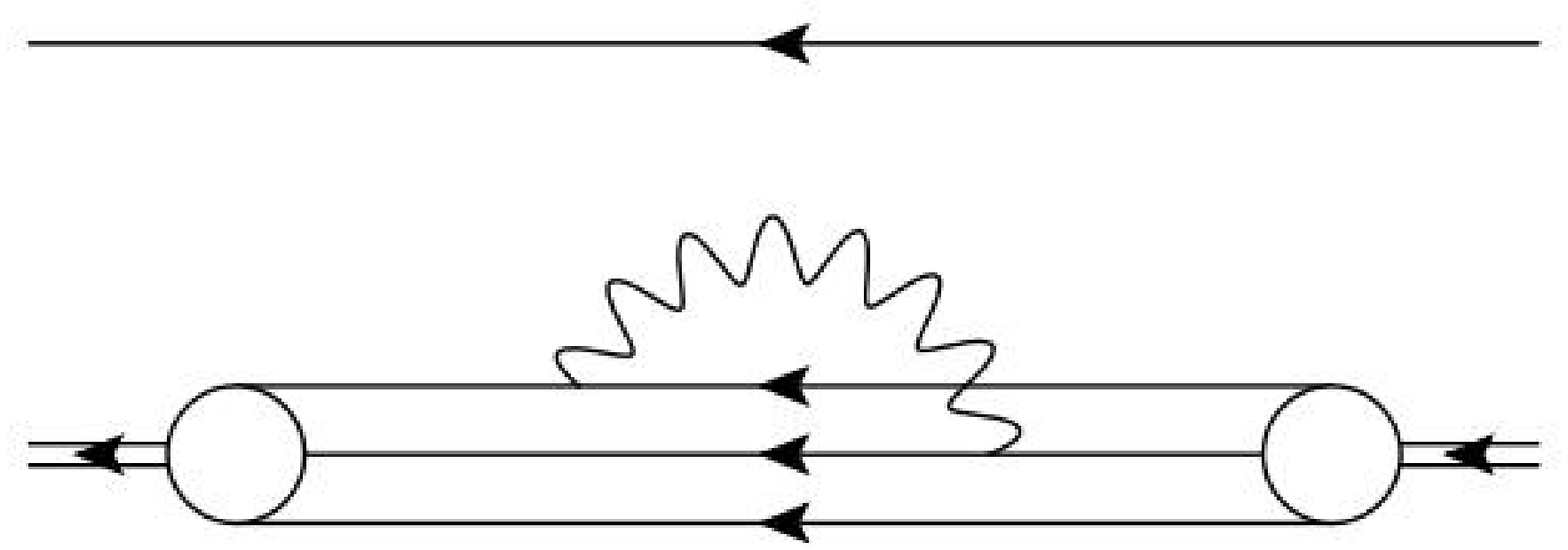}}
    \quad
    \subfigure[~Self-energy]
    {\label{qlamb2}\includegraphics[width=7cm]{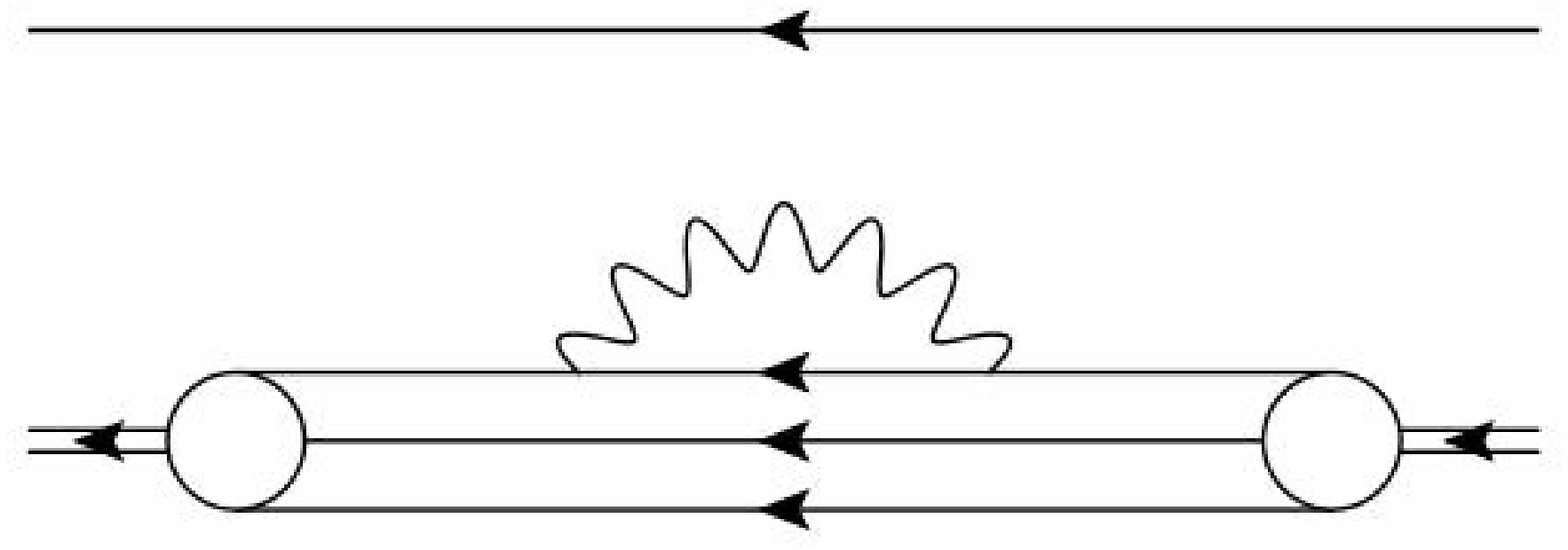}}
  \end{center}
  \caption{The two diagrams contributing to the Lamb shift in the proton 
    bound state.}
  \label{qlamb}
\end{figure}

The diagrams that involve one photon are given in
Figs.~\ref{Hlamb},~\ref{qlamb}, and \ref{1xqh}. Fig.~\ref{Hlamb} is the
standard one-loop Lamb shift, which has been evaluated to spectroscopic
accuracy. Fig.~\ref{qlamb} is the electromagnetic self energy of the
proton, which was calculated using the Feynman gauge in
Ref.~\cite{oldprl}. The simplest diagram to evaluate is
Fig.~\ref{qlamb1}, photon exchange between pairs of quarks. It
contributes $-0.222$ MeV to the proton electromagnetic self energy, all
of which comes from the vector part of the photon exchange. More
difficult to calculate is the self-energy diagram of Fig.~\ref{qlamb2}.
This diagram in general requires the inclusion of a self-mass
counter term, though not for the zero mass case. After this subtraction
an ultraviolet-divergent vertex term generally remains, but because the
quarks move freely within the well the ultraviolet divergent part of
this term vanishes, and numerical evaluation yields $0.658$ MeV (the
results given in Ref.~\cite{oldprl} referred to in the introduction are
based on the radius $R=1$ fm, but because they scale as $1/R$ a factor
$1.2$ must be inserted for comparison to the present work). We note that
vacuum polarization terms do not contribute for an isolated proton, but
may be of interest for muonic hydrogen, an issue that will be discussed
further in the conclusion. Finally, Figs.~3a and 3b describe exchange of
a photon between the lepton and the quarks in the proton and the
first-order effect of $H_\rX$ respectively, and we now turn to their
numerical evaluation.

\section{One-photon exchange}

\begin{figure}[h]
  \begin{center}
   \subfigure[~One photon exchange between the lepton and a quark]
    {\label{1xqh}\includegraphics[width=7cm]{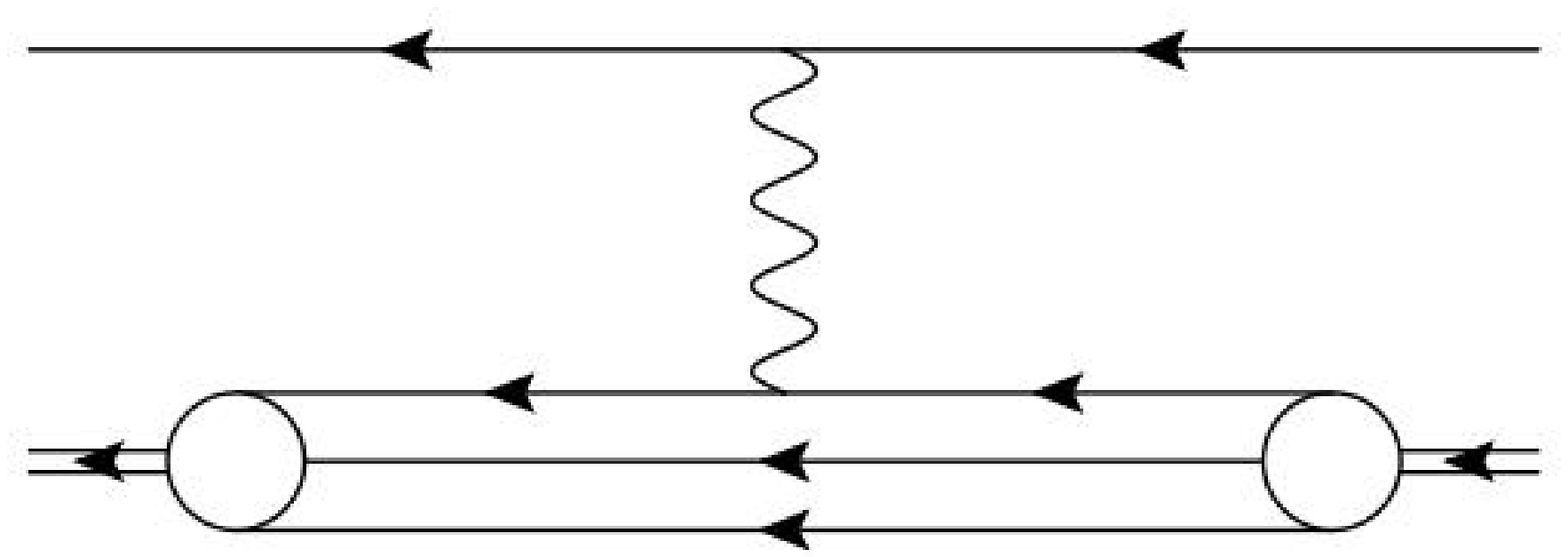}}  
   \qquad  
   \subfigure[~Counterterm contribution]
   {\label{ct}\includegraphics[width=7cm]{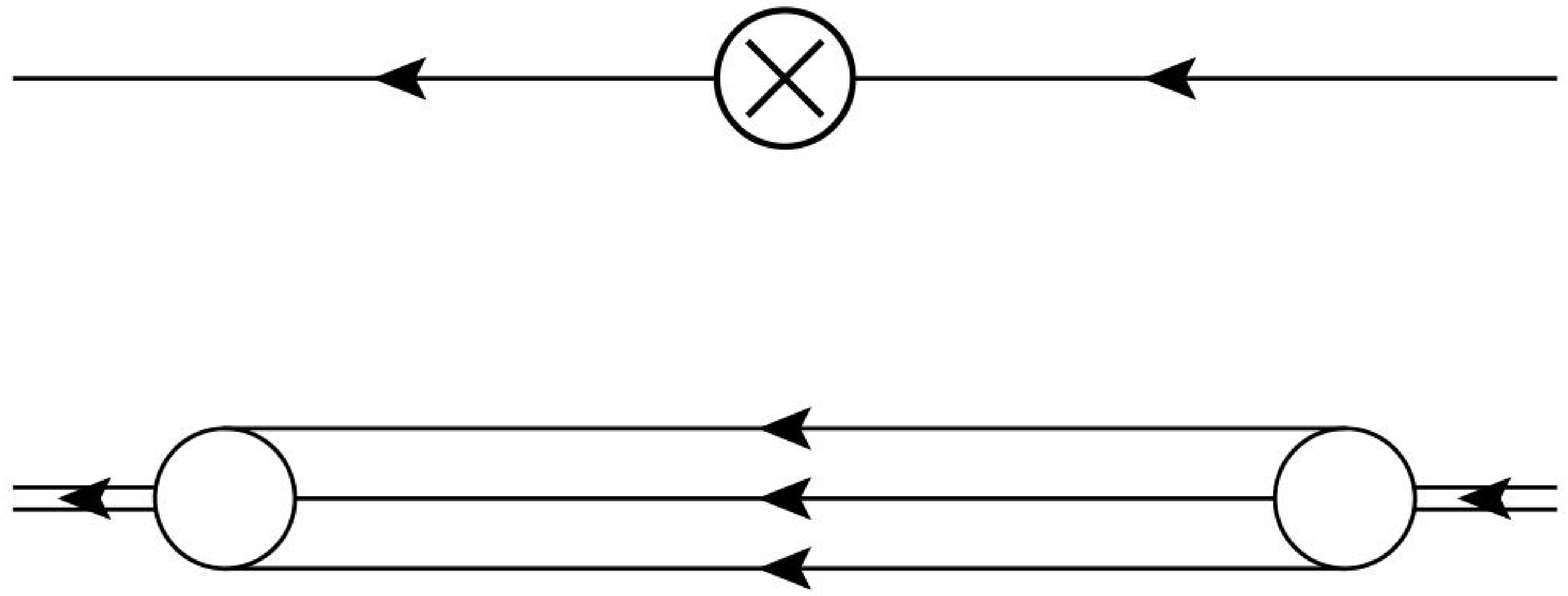}}
  \end{center}
  \caption{Order $\alpha$ contributions}
  \label{exchange1}
\end{figure}

We evaluate Fig.~\ref{1xqh} in the Coulomb gauge, and consider only Coulomb 
photon exchange. This leads to  the energy shift
\begin{equation}
 \Delta E_{\rm 1C}
 = -\alpha \int \frac{\rd\bm x \, \rd\bm y}{|\bm x - \bm y|} \,
    {\phi_v}^{\dagger}(\bm x) \phi_v(\bm x) \phi_\rg^{\dagger}(\bm
    y)\phi_\rg(\bm y).
\end{equation}
We have used our approximation of having the up and down quark wave
functions being equal, so the sum of the contribution of the three
quarks gives the single term $\phi_\rg^{\dagger} \phi_\rg$.  A direct
evaluation of this diagram for a state with principal quantum number $n$
gives a result very close to $-m_l\alpha^2/n^2$, with the difference
attributable to relativistic effects and the finite size of the proton
built into our model. The associated counter term in Fig.~\ref{ct}
contributes
\begin{equation}
 \Delta E_\rX 
 = \alpha \int \frac{\rd\bm x}{x} \,
 {\phi_v}^{\dagger}(\bm x) \phi_v(\bm x).
\end{equation}
For the $2s_{1/2}$ state it has the value
\begin{eqnarray}
 \Delta E_\rX & = & \frac{ m_l \alpha^2}{\gamma \sqrt{8(1+\gamma)}}\,,
\end{eqnarray}
where $\gamma \equiv \sqrt{1 - (Z \alpha)^2}$. 

Comparing the sum of $\Delta E_{1\rC}$ and $\Delta E_\rX$  to $E^{(1)}$
[Eq.~(\ref{firstorderpt})], one sees that it is exactly reproduced by
the field-theory expression.  However, because we now have a specific
model for $\rho(\bm x)$, we do not make the approximations made in the
QM treatment and instead numerically evaluate it. We find
\begin{eqnarray}
 E^{(1)}
 & = &   2.278\,68 \times 10^{-11}~ {\rm a.u.} ~~[\rm{hydrogen}] \nonumber \\
 E^{(1)}
 & = &   2.006\,26  \times 10^{-4}~~ {\rm a.u.} ~~[\rm{muonic~hydrogen}],
\end{eqnarray}
where all digits are significant.
The root-mean-square charge radius $r_{\rp}$ for $R=1.2$ fm is
\begin{eqnarray}
 r_{\rm p} & = & 0.728\,966 \, R \nonumber \\
     & = & 0.874\,760 \, {\rm fm} .
 \label{RvsRp}
\end{eqnarray}

We emphasize that this is not meant to be a prediction of the proton's
rms radius, it is simply the static-well result when $R=1.2$ fm.
However, if we use this in the standard nonrelativistic expression for
the finite size effect in hydrogen given in Eq.~(\ref{rms1}) we get
\begin{eqnarray}
 E^{(1)}_0
 & = &   2.277\,15 \times 10^{-11}~ {\rm a.u.} ~~[\rm{hydrogen}] \nonumber \\
 E^{(1)}_0
 & = &   2.013\,00 \times 10^{-4}~~ {\rm a.u.} ~~[\rm{muonic~hydrogen}],
\end{eqnarray}
which differ from the exact result by 0.07\,\% and 0.34\,\%
respectively.  Thus we see that the calculation reproduces the bulk of
the nonrelativistic expression for the effect of finite nuclear size on
$2s_{1/2}$ energy levels. In the following, while the basic parameter of
the static-well model is the well radius $R$, we use Eq.~(\ref{RvsRp})
to replace it by $1.3718 \, r_{\rp}$ in all formulas dependent on $R$.

As we have a complete model of the charge distribution, these small
deviations can be attributed to higher moments and relativistic effects.
By carrying out the calculation for a range of $R$ around 1.2 fm, we
find the fits
\begin{equation}
 E_{2s}^{(1)}(\tilde{r}_{\rp})
 = [2.977\,91 \times 10^{-11} \, \tilde{r}_{\rp}^{2\gamma} 
 - 6.188\,69 \times 10^{-16} \, \tilde{r}_{\rp}^3]\, {\rm a.u.}
\end{equation}
for hydrogen and
\begin{equation}
 E_{2s}^{(1)}(\tilde{r}_{\rp}) = [2.631\,70 \times 10^{-4} \, 
 \tilde{r}_{\rp}^{2\gamma} 
 - 1.126\,29 \times 10^{-6} \tilde{r}_{\rp}^3] \, {\rm a.u.}
\end{equation}
for muonic hydrogen, where $\tilde{r}_{\rp}$ denotes $r_{\rp}$ in units
of fermis (femtometers).

We first note that the coefficients of the first and second terms for
hydrogen increase by close to a factor of $(m_{\rmssmu}/m_\re)^3$ and
$(m_{\rmssmu}/m_\re)^4$ respectively for muonic hydrogen, consistent
with the dependence on $m_l$ shown in Eqns.~(\ref{rms1}) and
(\ref{rms2}). The coefficients agree at a level of the order of one
tenth of a percent.  We originally attempted the fit with a quadratic
term in $\tilde{r}_{\rp}$ instead of a term with the exponent $2\gamma$,
but were forced to use the latter form for hydrogen to get a proper fit.
(The effect is less important for muonic hydrogen).  In fact, it is
known that the actual dependence of leading finite-size correction on
$\tilde{r}_{\rp}$ is not the nonrelativistic quadratic form, but instead
the relativistic form used above, as shown in  Ref.~\cite{Mohrgamma}.
In perturbation theory, the leading effect of the fractional power is a
correction given by the Taylor expansion of $\tilde{r}_{\rm
p}^{2\gamma-2}$, which leads to a logarithmic term of relative order
$(Z\alpha)^2\ln{\tilde{r}_{\rp}}$.

\section{Screening of the proton self energy}
\label{sec:spse}

As mentioned in the introduction, the proton electromagnetic self energy
has contributions from the Feynman diagrams of Fig.~2.  Before
discussing how these diagrams are modified when the lepton interacts
with the proton, which can be thought of as the Lamb shift of the
proton, we mention another Lamb shift related term.  This other effect,
while negligible for hydrogen because it is of relative order
$(m_\rr/M_\rN)^2$, makes a small contribution for muonic hydrogen, and is
not suppressed at all for positronium,  accounting for the self energy
of the positron in that system. It was first derived by Fulton and
Martin \cite{Fulton-Martin}, and is given by
\begin{equation}
 E_{\rm SEN} = {4 \alpha^5 \over 3 \pi n^3} {m_\rr^3 \over m_\rN^2}  
 \left[ \ln \left({m_\rN \over m_\rr \alpha^2}\right) \delta_{l0} 
 - \ln\, k_0(n,l)\right].
\end{equation}
This recoil effect, which shifts the $2s_{1/2}$ energy in muonic
hydrogen by $0.010$ meV, is not included in our approach. 

\begin{figure}[h]
  \begin{center}
    \subfigure[~Proton Lamb shift vertex correction]
    {\label{pqls1}\includegraphics[width=7cm]{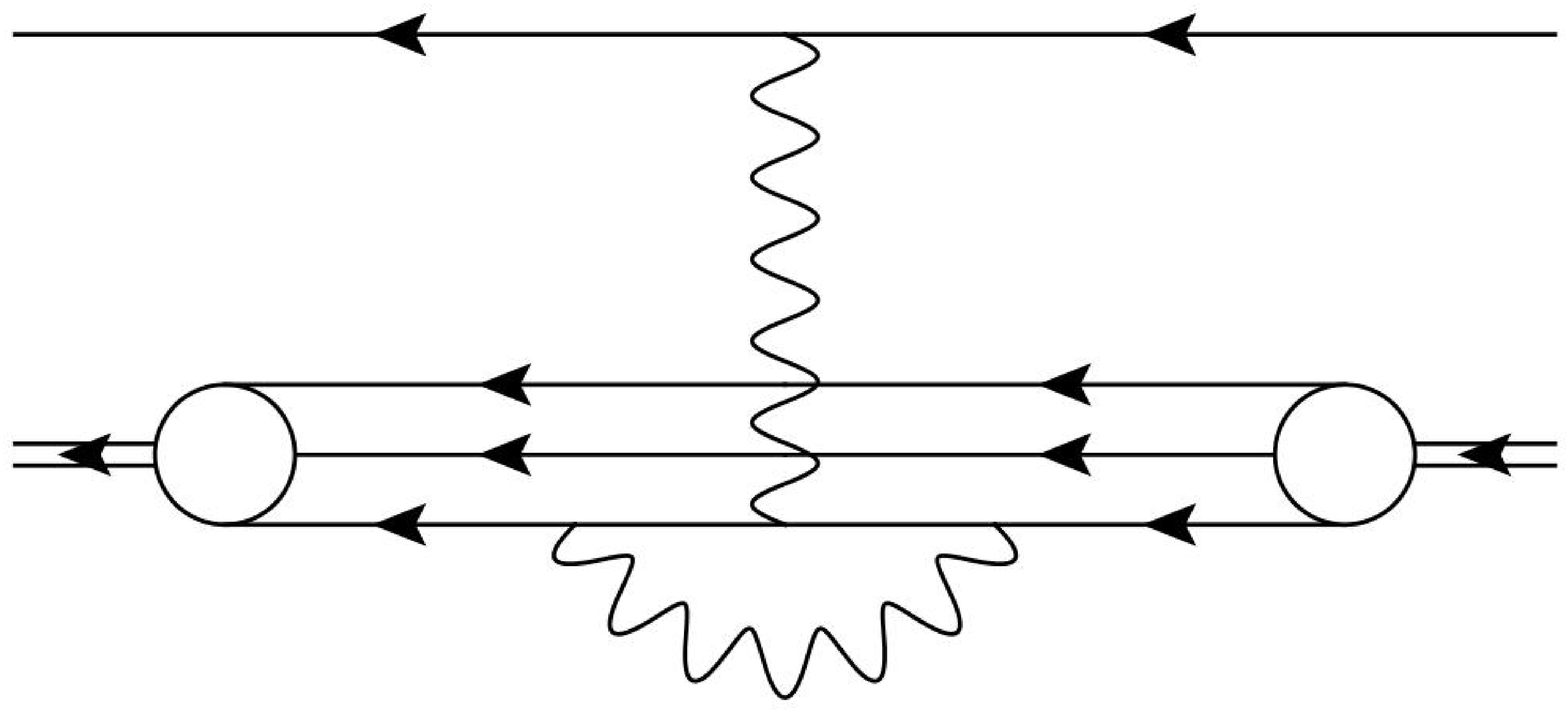}}
    \qquad  
    \subfigure[~Proton Lamb shift exchange correction]
    {\label{pqls2}\includegraphics[width=7cm]{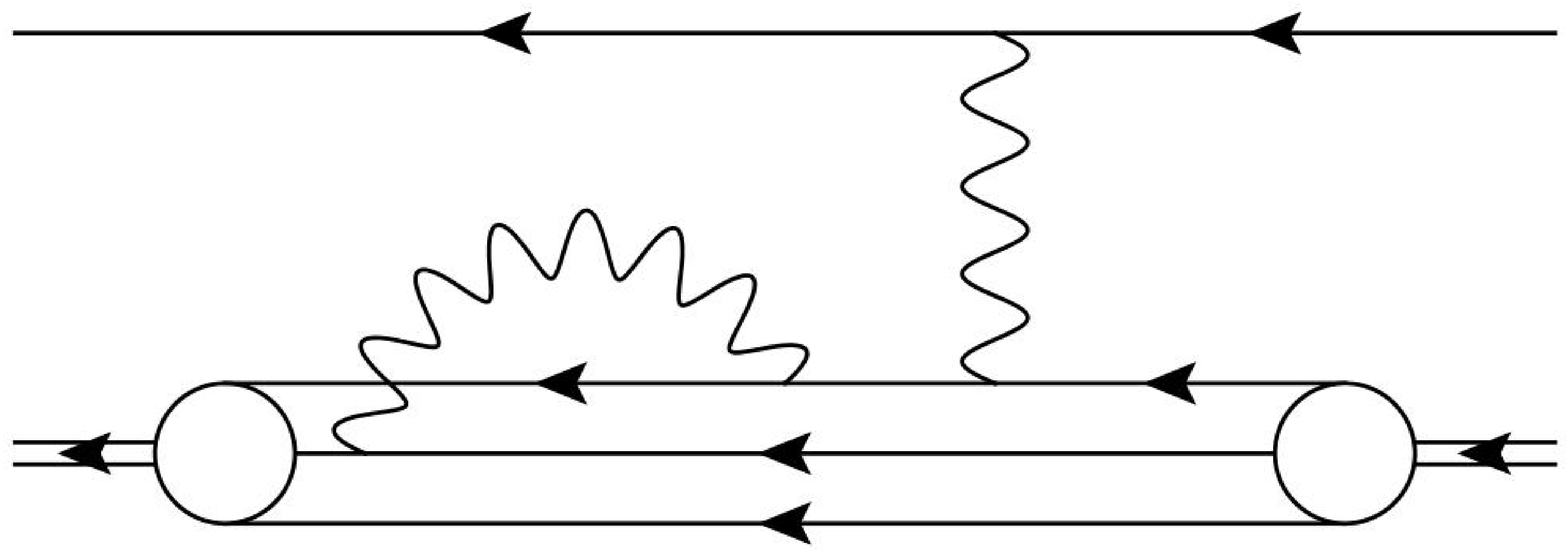}}
  \end{center}
  \label{exchange2}
  \caption{Order $\alpha^2$ corrections to proton Lamb shift}
\end{figure}

An isolated proton can of course emit and reabsorb a photon, giving rise
to the electromagnetic self energy of the proton just mentioned.  This
contributes to the mass of the proton, but when the proton is in a bound
state an additional shift arises, described in lowest order by the
Feynman diagrams in Figs.~\ref{pqls1} and \ref{pqls2}. We note in
passing that these diagrams do not have an analog in the QM treatment,
as they involve internal electromagnetic interactions in the proton. In
this paper we restrict our attention to the second of these diagrams,
which we refer to as exchange corrections to the EM self energy of the
proton and label as $\Delta E_{\rm ex}({\rm pLS})$.  This is justified
by the behavior of the lowest-order proton electromagnetic self energy,
where the size of the exchange term and the quark self energy are of the
same magnitude. The photon propagators are both taken to be Coulomb
propagators, and it is straightforward to show that this set of diagrams
gives the energy shift
\begin{eqnarray}
 \Delta E_{\rm ex}({\rm pLS}) 
 &=&- 2 \alpha^2 \,\frac{\tilde{q}_i \,\tilde{q}_j^2}{16 \pi^2} 
 \int \frac{\rd\bm x \,\rd\bm y \,\rd\bm z \,\rd\bm w} { | \bm x - \bm y|
   |\bm w- \bm z|} {\psi^{\dagger}}_v (\bm w) \psi_v (\bm w) 
   \nonumber\\&&\times
   \sum_n {1 \over \epsilon_\rg - \epsilon_n}
   \left< p \left| : \! {\psi^{\dagger}}_i(\bm x) \psi_i (\bm x) 
   {\psi^{\dagger}}_j(\bm y) \phi_n(\bm y) 
   \phi^{\dagger}_n(\bm z)\psi_j(\bm z) \! : \right|p \right> .
\end{eqnarray}
In this equation $i$ and $j$ are understood to be summed over the two
flavors and appropriate sums over color indices are implicit;
$\tilde{q}_i$ is the charge of the up or down quark in units of $e$
depending on whether $i=1$ or $2$ respectively; $\psi_i$, $\psi_j$, and
their adjoints are field operators, but $\phi_n$ and its adjoint are
wavefunctions, with the sum over $n$ going over all allowed values of
angular momentum, $\kappa_n$ and $\mu_n$, and the positive- and
negative-energy states associated with $\kappa_n$.  For two Coulomb
photons, one can show that only $\kappa_n=-1$ yields a non-zero
contribution.  After taking the normal-ordered product of the quark
fields between the spin up wave function of the proton and noting the
integrations over angles are all elementary, one is left with the sum  
\begin{eqnarray}
 \Delta E_{\rm ex}({\rm pLS}) 
 = \frac{8}{9}\, \alpha^2 \sum_{[n]} \frac{1}{\epsilon_\rg - \epsilon_n} 
 \int_0^R \rd x \,x^2 R_{\rg\rg}(x) \int_0^R \rd y \,y^2 R_{\rg n}(y)  
 \nonumber \\
 \int_0^R \rd z \,z^2 R_{n\rg}(z) \int_0^{\infty} \rd w \,w^2 R_{vv}(w)
 \,\frac{1}{{\max}(x,y)} \, \frac{1}{{\max}(z,w)}\,,
 \label{eq:expls}
\end{eqnarray}
where $[n]$ denotes summation over all $s$-states except the $1s$ state,
and
\begin{equation}
R_{ij}(x) \equiv  {f_1}_i(x) {f_1}_j(x) + {f_2}_i(x) {f_2}_j(x).
\label{eq:sprod}
\end{equation}
We find, using finite basis sets to carry out the sum over $[n]$, that
\begin{equation}
 \Delta E_{\rm ex}{\rm(pLS)} =  0.5 \times 10^{-3} ~{\rm meV}.
\end{equation}
The smallness of this effect is related to the fact the integral over $z$ in 
Eq.~(\ref{eq:expls}) is
\begin{equation}
 \int_0^R \rd z \,z^2 R_{n\rg}(z) \,\frac{1}{\max(z,w)}\,,
\end{equation}
which vanishes when $w>R$ from the orthogonality of the radial wave
functions.  This restricts the muon wave function to lie within the
nucleus, which gives a large suppression factor that scales as
$(R/a_0)^3$.

This same suppression factor should make the diagram of 4b, which is
more difficult to evaluate, numerically unimportant.  However, an
interesting application of our approach would be the calculation of
vacuum polarization. This was shown in Ref. \cite{oldprl} to vanish for
a free proton, but when bound in an atom the arguments for its vanishing
no longer apply, and in fact this should correspond to the effect of
hadronic vacuum polarization, which is treated as a small effect,
estimated in Ref. \cite{Pohl} to be 0.011 meV.  However, we are not
aware of any direct calculation of this term with zero-mass quarks, and
will discuss how such a calculation could be carried out in the
conclusions section.

Before turning to two-photon effects, we give more details about the use
of finite basis sets. As described in Ref.~\cite{bsplines}, atomic
finite basis set  calculations are carried out in a well much larger
than the atom, but with the same boundary conditions as used for the
quarks. We continue to use an atomic basis set appropriate for hydrogen,
but add a second basis set for the quarks, which is obtained by simple
modifications of the atomic code, involving changing the fermion mass to
zero, eliminating the potential, and changing from atomic units to
MeV-fm units. Atomic grids are created on an exponential grid of the
form $r(i) = r_0 \left[e^{h(i-1)}-1\right]$. We use the same grid for
both quark and lepton wavefunctions. This is done by choosing parameters
such that if, for example, a 1000 point grid were used for the atom, the
two hundredth point would be at $r=1.2$ fm, so that the quark wave
function would be put on a 200 point grid that matched the atomic grid,
though of course the quark wave function vanishes for $i > 200$.
Several grids were used to test numerical stability.  As is also the
case for leptons, a complete set of positive and negative energy states
result for each possible value of $\kappa_n$, in this case $N$ positive
energy $s$-states and $N$ negative energy $s$-states, with a typical
value of $N$ being 50. For leptons the effect of the negative-energy
states is generally very small, entering at the order of the Lamb shift.
However, for quarks they play a more important role.

\section{Two-photon exchange}

We now turn our attention to diagrams shown in Fig.~5. In this section
we will show that they in part reproduce the second order perturbation
theory expression for $E^{(2)}$, Eq.~(\ref{secondorderpt}) in Section
II,  but have in addition extra terms we identify as proton
polarizability.  To compare with individual diagrams, it is useful to
employ Eq.~(\ref{Vparts}) to represent $\delta V$ in terms of $\rho -
\delta$, which yields four terms for Eq.~(\ref{secondorderpt}):
\begin{equation}
E^{(2)\,}_{\delta \delta} = \alpha^2 
\int \rd\bm x \, \rd\bm y \,
\phi_v^\dagger(\bm x)\, \frac{1}{|\bm x|}
\sum_{m\ne v} 
\frac{\phi_m(\bm x)\,\,\phi_m^{\dagger}(\bm y)}
{\epsilon_v - \epsilon_m} \,
\frac{1}{|\bm y|} \,\phi_v(\bm y),
\label{pt11}
\end{equation}
\begin{equation}
E^{(2)\,}_{\delta \rho} = -\alpha^2 
\int \rd\bm x \, \rd\bm y \, \rd\bm w \,
\phi_v^\dagger(\bm x)\, \frac{1}{|\bm x|} 
\sum_{m\ne v} 
\frac{\phi_m(\bm x)\,\,\phi_m^{\dagger}(\bm y)}
{\epsilon_v - \epsilon_m} \,
\frac{\rho(\bm w)\,}{|\bm y- \bm w|} \, \phi_v(\bm y),
\label{pt12}
\end{equation}
\begin{equation}
E^{(2)\,}_{\rho \delta} = -\alpha^2
\int \rd\bm x \, \rd\bm y \, \rd\bm z \,
\phi_v^\dagger(\bm x)\, \frac{\rho(\bm z)}{|\bm x- \bm z|} 
\sum_{m\ne v} 
\frac{\phi_m(\bm x)\,\,\phi_m^{\dagger}(\bm y)}
{\epsilon_v - \epsilon_m} \,
\frac{1}{|\bm y|} \, \phi_v(\bm y),
\label{pt21}
\end{equation}
and
\begin{equation}
E^{(2)\,}_{\rho \rho} = \alpha^2
\int \rd\bm x \, \rd\bm y \, \rd\bm z \, \rd\bm w \,
\phi_v^\dagger(\bm x)\, \frac{\rho(\bm z)}{|\bm x - \bm z|} 
\sum_{m\ne v} 
\frac{\phi_m(\bm x)\,\,\phi_m^{\dagger}(\bm y)}
{\epsilon_v - \epsilon_m} \,
\frac{\rho(\bm w)\,}{|\bm y- \bm w|} \, \phi_v(\bm y).
\label{pt22}
\end{equation}

The simplest diagram, Fig. 5a, is easily shown to give $E^{(2)}_{\delta
\delta}$. The diagrams of Fig. 5b and its complex conjugate give
$E^{(2)}_{\delta \rho}$ and $E^{(2)}_{\rho \delta}$ if we identify
$\rho(\vec x) = \rho_\rg(\vec x)$ as in the treatment of one-photon
exchange. This leaves only $E^{(2)}_{\rho \rho}$ to be accounted for. In
general one contribution to it comes from Fig. 5c, given by
\begin{eqnarray}
\Delta E_2(ij) &=& \alpha^2 \tilde{q}_i \, \tilde{q}_j 
\sum_m \int \frac{\rd\bm x \, \rd\bm y \, \rd\bm z \, \rd\bm w}
{|\bm x - \bm z| | \bm y - \bm w|} \,
\frac{\phi_v^{\dagger}(\bm x) \phi_m(\bm x) 
\phi_m^{\dagger}(\bm y) \phi_v(\bm y)}{\epsilon_v-\epsilon_m} 
  \nonumber\\&&\times
  \left< p \left| :\psi^{\dagger}_i(\bm z) \psi_i(\bm z) 
  \psi^{\dagger}_j(\bm w) \psi_j(\bm w) : \right| p \right>,
\end{eqnarray}
where the definition of $\tilde{q}_i$ is the same as in
Sec.~\ref{sec:spse}.  The various contributions to this term happen to
sum to zero for the proton in our model, but we note they would not were
we considering the neutron, or if we were taking the up and down quark
wave functions to be different. 

The final two diagrams, 5d and 5e, are referred to as the ladder (L) and
crossed ladder (XL) respectively.  The closed loop in these diagrams is
associated with  an integration over a virtual energy.  Because in this
paper we consider only two-Coulomb photon exchange, the analysis of the
loop integral is considerably simpler than the case where the photon
propagators have energy dependence. That complication was encountered in
the full Feynman gauge analysis of ladder and crossed ladder diagrams in
excited states of heliumlike ions \cite{heliumexcited}, upon which the
present calculation is patterned. Issues involved in carrying out the
full calculation will be discussed in the conclusion. In this simpler
case we can carry out the integral over the timelike component of the
loop momentum with Cauchy's theorem, which requires identifying the
poles coming from the propagators. We partition the lepton and quark
propagators into positive and negative energy parts, which could lead to
four contributions for each diagram, but the position of the poles is
such that only two contributions fail to vanish. The surviving terms are
\begin{eqnarray}
\Delta E_\rL(++) & = & \tilde{q_i}^2 \alpha^2 
\int \frac{\rd\bm x \, \rd\bm y \, \rd\bm z \, \rd\bm w}
{|\bm x - \bm z||\bm y -\bm w|}
\sum_{m_+ n_+}\frac{\phi^{\dagger}_v(\bm x)
\phi_m(\bm x) \phi^{\dagger}_m(\bm y) \phi_v(\bm y)}
{\epsilon_v + \epsilon_\rg - \epsilon_m - \epsilon_n}  
\nonumber \\
&&\left< p\left|:\!\psi^{\dagger}_{q_i}(\bm z) 
\psi_n(\bm z) \psi^{\dagger}_n(\bm w) 
\psi_{q_i}(\bm w)\!:\right|p \right>
\end{eqnarray}
and
\begin{eqnarray}
\Delta E_\rL(--) & = & - \tilde{q_i}^2 \alpha^2 
\int \frac{\rd\bm x \, \rd\bm y \, \rd\bm z \, \rd\bm w}
{|\bm x - \bm z||\bm y -\bm w|}
\sum_{m_- n_-}\frac{\phi^{\dagger}_v(\bm x)
\phi_m(\bm x) \phi^{\dagger}_m(\bm y) \phi_v(\bm y)}
{\epsilon_v + \epsilon_\rg - \epsilon_m - \epsilon_n}  
\nonumber \\
&&\left< p\left|:\!\psi^{\dagger}_{q_i}(\bm z) 
\psi_n(\bm z) \psi^{\dagger}_n(\bm w) 
\psi_{q_i}(\bm w)\!:\right|p \right>
\end{eqnarray}
for the ladder diagram, and
\begin{eqnarray}
\Delta E_\rXL(+-) & = & \tilde{q_i}^2 \alpha^2 
\int \frac{\rd\bm x \, \rd\bm y \, \rd\bm z \, \rd\bm w}
{|\bm x - \bm w||\bm y -\bm z|}
\sum_{m_+ n_+}\frac{\phi^{\dagger}_v(\bm x)
\phi_m(\bm x) \phi^{\dagger}_m(\bm y) \phi_v(\bm y)}
{\epsilon_\rg - \epsilon_v + \epsilon_m - \epsilon_n}
\nonumber \\
&&\left< p\left|:\!\psi^{\dagger}_{q_i}(\bm z) 
\psi_n(\bm z) \psi^{\dagger}_n(\bm w) 
\psi_{q_i}(\bm w)\!:\right|p \right>
\end{eqnarray}
and
\begin{eqnarray}
\Delta E_\rXL(-+) & = & - \tilde{q_i}^2 \alpha^2 
\int \frac{\rd\bm x \, \rd\bm y \, \rd\bm z \, \rd\bm w}
{|\bm x - \bm w||\bm y -\bm z|}
\sum_{m_- n_-}\frac{\phi^{\dagger}_v(\bm x)
\phi_m(\bm x) \phi^{\dagger}_m(\bm y) \phi_v(\bm y)}
{\epsilon_\rg - \epsilon_v + \epsilon_m - \epsilon_n}
\nonumber \\
&&\left< p\left|:\!\psi^{\dagger}_{q_i}(\bm z) 
\psi_n(\bm z) \psi^{\dagger}_n(\bm w) 
\psi_{q_i}(\bm w)\!:\right|p \right>
\end{eqnarray}
for the crossed ladder. The relative minus sign between the two ladder
and the two crossed ladder terms comes from closing the contour in
different ways, and we note that the Coulomb propagators are different
for the ladder and crossed ladder diagrams. In the following we will
give formulas only for $\Delta E_\rL(++)$ and $\Delta E_\rXL(-+)$ as the
other terms differ from these only by an overall minus sign and a
reversed role of positive and negative energy states.

\subsection{Elastic contribution}

At this point we divide the sum over intermediate quark states into a
part with $n=\rg$, which leaves the proton unchanged and corresponds to
elastic scattering, and the remaining part, corresponding to inelastic
scattering, which gives the polarizability contribution. We now show
that the elastic term contains $E^{(2)}_{\rho \rho}$. As $\epsilon_n$ is
positive energy, only $\Delta E_\rL(++)$ and $\Delta E_\rXL(-+)$
contribute.  Because the quark states are now all $s$-states, the
angular-momentum dependence is particularly simple. The Coulomb
propagators can be simplified with the replacements
\begin{eqnarray}
\frac{1}{|\bm x - \bm z|} \rightarrow \frac{1}{r_{1a}}
\nonumber \\
\frac{1}{|\bm y - \bm w|} \rightarrow \frac{1}{r_{2a}} 
\end{eqnarray}
with $r_{1a} = \max(|\bm x|, |\bm z|)$, 
$r_{2a} = \max(|\bm y|,|\bm w|)$. 
(For later use we define 
$r_{1b} = \min(|\bm x|, |\bm z|)$, 
$r_{2b} = \min(|\bm y|,|\bm w|)$.)
Integration over angles then yields
\begin{eqnarray}
\Delta E_L(++) = \alpha^2 \int \rd V 
\frac{1}{r_{1a} r_{2a}} \sum_{m_+}
\frac{R_{vm}(x) R_{mv}(y) R_{\rg\rg}(z) R_{\rg\rg}(w)}
{\epsilon_v - \epsilon_m },
\end{eqnarray}
with $m$ being forced to be an $s$ state.  We have introduced the
shorthand
\begin{equation}
\int \rd V \equiv \int_0^{\infty}  \rd x \, x^2
\int_0^{\infty}  \rd y \, y^2 \int_0^R \rd z \, z^2
\int_0^R \rd w \, w^2.
\end{equation}

Because $R_{\rg\rg}(w)/4\pi = \rho_\rg(w)$, this reproduces the part of
$E^{(2)}_{\rho \rho}$ in which the electron propagator involves sums
over positive energy states. In order to include the part with negative
energy states the crossed ladder must be considered. The minus sign
mentioned above is needed for the energy denominators to match, and the
fact that $\rho(\bm z) \rho(\bm w)$ is symmetric under interchange of
$\bm z $ and $\bm w$ is also needed. Thus part of the ladder and crossed
ladder diagrams together with the other diagrams of Fig.~5 simply
reproduce second-order perturbation theory with a particular form for
the charge distribution of the proton.

We evaluated these terms numerically as a test of the coding, since the
QM treatment yields a nonrelativistic (NR) limit for comparison.  The
result for muonic hydrogen is
\begin{equation}
E_{2s}^{(2)} = -3.948 \times 10^{-7} ~{\rm a.u.},
\end{equation}
which is within $4\,\%$ of the NR formula. We recall that formula has a
mixture of the Zemach term and a $\langle r^3 \rangle$ term, where the
latter cancels part of the one-loop result.

\subsection{Proton polarizability}

We now turn to the evaluation of the remaining parts of the ladder and
crossed ladder. The quark propagator is treated in  the same manner as
described in Sec.~\ref{sec:spse}.  The feature of the basis set
mentioned in that section, whereby it automatically breaks into a
positive and a negative energy part, makes separation of the various
diagrams a simple matter. We begin with the case where $\kappa_m =
\kappa_n = -1$, but where we do not allow the intermediate quark to be
in the ground state. The result for the ladder is $0.000\,5$ meV, and is
almost completely cancelled by the crossed ladder, which contributes
$-0.000\,5$ meV, making this channel completely negligible.

In order to treat higher partial waves we must now specify the spin
state of the atom. Polarizability is important only for the $2s_{1/2}$
state, which of course can be a triplet or singlet state once the spin
of the proton is considered.  For one-Coulomb-photon exchange these
states have the same energy, as hyperfine splitting comes from
transverse photon exchange, which is not treated here. However, with
two-photon exchange, even when both are Coulomb photons, the triplet and
singlet energies differ except for the $l=0$ partial wave discussed
above. We therefore present in the following formulas for the energy
levels of the $2s_{1/2}$ singlet and triplet.

While each Coulomb propagator has its own partial wave expansion,
leading in general to a double sum over $l_1$ and $l_2$, the fact that
the quarks are all in $s$-states leads to the simplification that these
values are equal, $l_1=l_2\equiv l$.  Another simplification is that for
a given $l$, only $\kappa_m$ and $\kappa_n$ values associated with that
value give a non-vanishing contribution. We define angular factors
$A_l(\kappa_m,\kappa_n)$ for the ladder and $B_l(\kappa_m, \kappa_n)$
for the crossed ladder diagrams, which are rational fractions resulting
from the integrations over angles and sums over magnetic quantum numbers,
tabulated in Table II for $l=1$ and $l=2$. Both the triplet and singlet
state results are given, though in this paper we are only interested in
fine structure.  $l=1$ corresponds to the dominant dipole transition.
In terms of these coefficients, the formula for the ladder is
\begin{eqnarray}
\Delta E_L(++) = \alpha^2 \sum_l \int \rd V \,
 \frac{r_{1b}^l  r_{2b}^l}{r_{1a}^{l+1} r_{2a}^{l+1}} 
 \sum_{m_+ n_+ }A_l(\kappa_m,\kappa_n) \,
 \frac{R_{vm}(x) R_{mv}(y) R_{\rg n}(z) R_{n\rg}(w)}
 {\epsilon_v +\epsilon_\rg+\epsilon_n - \epsilon_m }
\end{eqnarray}
and for the crossed ladder is
\begin{eqnarray}
 \Delta E_X(+-) = \alpha^2 \sum_l \int \rd V \,
 \frac{r_{3b}^l  r_{4b}^l}{r_{3a}^{l+1} r_{4a}^{l+1}} 
 \sum_{m_+ n_-}B_l(\kappa_m,\kappa_n) \,
 \frac{R_{vm}(x) R_{mv}(y) R_{\rg n}(z) R_{n\rg}(w)}
 {\epsilon_v - \epsilon_\rg + \epsilon_n - \epsilon_m },
\end{eqnarray}
where $r_{3a} = {\max}(|\bm x|, |\bm w|), r_{3b} 
= {\min}(|\bm x|,|\bm w|)$
and $r_{4a} = {\max}(|\bm y|, |\bm z|), r_{4b} 
= {\min}(|\bm y|,|\bm z|)$.

~

\begin{table}[h]
\hrule\vspace{2 pt} \hrule
  \subtable{
    \begin{tabular}{c c<{\hspace{0.2cm}} c c c c}
      \multicolumn{6}{c}{$l=1$}  \\ 
      $\kappa_m$ & $\kappa_n$ & $A_1$(T) & 
                   $A_1$(S) & $B_1$(T) & $B_1$(S) \\ \hline
       1 &  1 & ~17/729 & 57/729  & 37/729  & 17/729\\
       1 & -2 & ~64/729 & 24/729  & 44/729  & 64/729 \\
      -2 &  1 & ~64/729 & 24/729  & 44/729  & 64/729 \\
      -2 & -2 & ~98/729 & 138/729 & 118/729 & 98/729 \\
    \end{tabular}
  }\hspace{10 pt}\vline\hspace{5 pt}
  \subtable{
    \begin{tabular}{c c<{\hspace{0.2cm}} c c c c}
      \multicolumn{6}{c}{$l=2$}  \\ 
      $\,\kappa_m$ & $\kappa_n$ & $A_2$(T) & 
                   $A_2$(S) & $B_2$(T) & $B_2$(S) \\ \hline
       2 &  2 & ~26/1125 & 66/1125  & 46/1125 & 26/1125\\
       2 & -3 & ~64/1125 & 24/1125  & 44/1125 & 64/1125 \\
      -3 &  2 & ~64/1125 & 24/1125  & 44/1125 & 64/1125 \\
      -3 & -3 & ~71/1125 & 111/1125 & 91/1125 & 71/1125 \\
    \end{tabular}
  }
\hrule\vspace{2 pt} \hrule
\caption{Angular coefficient factors for the allowed values of $\kappa_m$, 
$\kappa_n$ for $l=1$ and $l=2$.  Factors of $A_l$ are for the ladder 
diagram;  factors of $B_l$ are for the crossed ladder.  ``T'' refers to the 
triplet atomic state, while ``S'' refers to the singlet.}
\label{tab:clebsches}
\end{table}
 ~

These can be evaluated with the techniques described above with only
simple changes, as the spline basis set contains all values of $\kappa$
needed and the radial integrals over $r_b^l/r_a^{l+1}$ are evaluated
with techniques valid for arbitrary $l$. The main numerical problem was
ensuring that the typical rapid convergence of splines in atomic physics
carried over to this problem. The calculation shows that the effect of
the $l=2$ channel is very small, with almost the entire effect of
polarizability coming from the $l=1$ dipole channel, which shifts the
$2s_{1/2}$ energy down by 0.026 meV.  This leads to our main
polarizability result for the splitting,
\begin{equation}
E(2s_{1/2}-2p_{3/2}) = 0 .026 \, \mbox{meV},
\end{equation}
consistent with the results from dispersion relation analyses, as will
be discussed in the conclusions.

\section{Static polarizabilities}

One of the basic electromagnetic properties of the proton is its static
polarizability, which has an electric and a magnetic component. We
concentrate on the static electric polarizability, which is  given by
the Particle Data Group as
\begin{equation}
\alpha_{\rm p}  =  12.0(6)\, \times10^{-4}  \mbox{ fm}^3. 
\label{eq:epol}
\end{equation}
While this value is extracted from Compton scattering data together with
dispersion theory arguments \cite{Lvov}, in the static-well model the
lowest-order expression comes from the diagram of Fig.~6, where the line
with a cross indicates a constant electric field with magnitude $E_{\rm
ext}$.  The energy shift is related to the static electric
polarizability through
\begin{equation}
\Delta E_\alpha = -\frac{1}{2}\, \alpha_{\rm p}\,E_{\rm ext}^2.
\end{equation}
We note two similar calculations using the closely related MIT bag model
\cite{Schafer}, \cite{Bertsch} have been presented, but detailed
comparison with their results is not possible. A factor of $4 \pi$ has
been absorbed into the external field.

The basic equation for the energy shift of a proton in the presence of a
constant electric field described by the potential $\phi(\bm r) =
-E_{\rm ext} x_3$ is
\begin{equation}
\Delta E_{\alpha}  =  \tilde{q_i}^2 E_{\rm ext}^2
\sum_{n}\frac{1}{\epsilon_\rg - \epsilon_n}  
\int \rd\bm x \, \rd\bm y 
\left< p\left|:\!\psi^{\dagger}_{q_i}(\bm x) x_3\psi_n(\bm x) 
\psi^{\dagger}_n(\bm y)y_3 \psi_{q_i}(\bm y)\!:\right|p \right>
\end{equation}
which reduces to
\begin{equation}
\Delta E_{\alpha}   =   \frac{\alpha}{9} E_{\rm ext}^2  \sum_{n}  
\int \rd\bm x \, \rd\bm y  \,
\frac{ 7 \phi^{\dagger}_a(\bm x) x_3\phi_n(\bm x) 
\phi^{\dagger}_n(\bm y)y_3 \phi_a(\bm y) + 
2 \phi^{\dagger}_b(\bm x) x_3\phi_n(\bm x) 
\phi^{\dagger}_n(\bm y)y_3 \phi_b(\bm y)}{\epsilon_\rg- \epsilon_n} .
\qquad
\end{equation}
For this case the orientation of the quark spin does not matter, so the
expression simplifies to 
\begin{equation}
\alpha_{\rm p}  = -2 \sum_{n}\frac{1}{\epsilon_\rg - \epsilon_n}  
\int \rd\bm x \, \rd\bm y  \,
\phi^{\dagger}_\rg(\bm x) x_3\phi_n(\bm x) \phi^{\dagger}_n(\bm y)y_3
\phi_\rg(\bm y),
\label{alphapdef}
\end{equation}
where we have now isolated the factor $-\frac{1}{2} E_{\rm ext}^2$ to
give the formula in terms of $\alpha_{\rm p}$.  We emphasize at this
point that this is a field theoretic derivation, and that the sum over
$n$ is complete, including negative-energy states.

Introducing the notation [see Eq.~(\ref{eq:sprod})]
\begin{equation}
r_{ij} \equiv \int_0^R \rd x \, x^3 \, R_{ij}(x),
\end{equation} 
after integration over coordinate angles we find
\begin{equation} 
\alpha_\rp =-\frac{2}{9} \sum_{n\atop \kappa_n=1}\,\frac{r_{\rg n} r_{n\rg}}
{\epsilon_\rg-\epsilon_n}  -
\frac{4}{9} \sum_{n\atop \kappa_n=-2}\,\frac{r_{\rg n}r_{n\rg}}
{\epsilon_\rg-\epsilon_n}. 
\label{eq:polsumnew}
\end{equation}

We calculated the two terms with basis set techniques. The most striking
result found was that the $\kappa=1$ channel vanishes.  If one sums over
only positive energy states one obtains a nonzero result, but inclusion
of the negative energy states leads to an exact cancellation.  This
situation does not occur for the $\kappa=-2$ channel, which in our model
is solely responsible for the proton static polarizability. 

We also calculated $\alpha_{\rm p}$ using a form for the quark
propagator that does not rely on the spectral decomposition given in
Eq.~(\ref{quarkprop}). After rewriting Eq.~(\ref{alphapdef}) as
\begin{equation}
\alpha_p = 2 \int \rd\bm x \, \rd\bm y  \,
\phi_\rg^{\dagger}(\bm x) x_3 G_\rq(\bm x, \bm y; \epsilon_\rg) 
y_3 \phi_\rg(\bm y) ,
\label{eq:spol}
\end{equation}
we use the fact that the quark Green function satisfies the differential
equation
\begin{equation}
(-\ri\bm \alpha \cdot \bm \nabla_{\!\bm x} + \beta m_\rq - \epsilon_\rg)
G_\rq(\bm x, \bm y;\epsilon_\rg) = \delta^3( \bm x - \bm y)
\end{equation}
with appropriate boundary conditions. Without those conditions this is
the same equation that a free fermion Green function satisfies, and a
simple modification of the well-known partial-wave expansion of the
latter propagator can be made to solve for the quark propagator. We
illustrate this for the Green function of a massless quark with positive
$\kappa=l$ and $x > y$, in which case
\begin{eqnarray}
G_\rq(\bm x, \bm y;\epsilon_\rg) &=& 
i \epsilon_g^2 
\sum_{\kappa \mu}  
\left( \begin{array}{c}  \left[h_l(\epsilon_\rg x) 
+ A_l j_l(\epsilon_\rg x)\right] \chi_{\kappa \mu}(\bm{\hat{x}}) \\
\ri \left[h_{l-1}(\epsilon_\rg x) + A_l j_{l-1}(\epsilon_\rg x)\right]
\chi_{-\kappa \mu}(\bm{\hat{x}}) \end{array} \right) 
\nonumber\\[10 pt]&&\qquad\qquad\qquad\times
\left(\begin{array}{c@{\qquad}c}
j_l(\epsilon_\rg y) \chi^{\dagger}_{\kappa \mu}(\bm{\hat{y}})
& -\ri j_{l-1}(\epsilon_\rg y)
\chi^{\dagger}_{-\kappa \mu}(\bm{\hat{y}})\end{array}\right).
\end{eqnarray}
The coefficients $A_l$ are determined by the boundary condition. The
fact that the propagator ranges over a finite volume allows the
admixture of the solution proportional to $A_l$ which is forbidden for a
free propagator because of the boundary condition as
$x\rightarrow\infty$.  When $\kappa=1$, the integrand in
Eq.~(\ref{eq:spol}) includes the factor
\begin{eqnarray}
\left(\begin{array}{c@{\qquad}c}
j_1(\epsilon_\rg y) \chi^{\dagger}_{1 \mu}(\bm{\hat{y}})
& -\ri j_0(\epsilon_\rg y)
\chi^{\dagger}_{-1 \mu}(\bm{\hat{y}})\end{array}\right)
\left(\begin{array}{c}
j_0(\epsilon_\rg y) \chi^{\dagger}_{-1 \mu}(\bm{\hat{y}})
\\ -\ri j_1(\epsilon_\rg y)
\chi^{\dagger}_{1 \mu}(\bm{\hat{y}})\end{array}\right),
\end{eqnarray}
which vanishes.  An analogous factor on the left-hand-side when $x<y$
also vanishes.  However, for $\kappa=-2$ there is no such cancellation,
and using
\begin{equation}
A_1 = \frac{h_2(\epsilon_\rg R)-h_1(\epsilon_\rg R)}
{j_2(\epsilon_\rg R)-j_1(\epsilon_\rg R)},
\end{equation} 
we find the numerical result
\begin{equation}
\alpha_\rp = 25.4 \times 10^{-4} \mbox{ fm}^3 \, .
\end{equation}

The spectral decomposition discussed previously gives the same result,
and one can see the sum is dominated by the first $p_{3/2}$ state, with
higher-$n$ positive-energy and negative-energy states entering at under
one tenth of a percent.  This result is a factor of two larger than the
experimental value in Eq.~(\ref{eq:epol}). However, we note that the
relation of the PDG result quoted to static polarizability determined
from energy shifts involves some subtleties, discussed in \cite{Lvov}
and more recently in \cite{newHill}.  

\section{Conclusions}

In this paper we present an approach to calculating the effect of the
electromagnetic structure of the proton on energy levels of muonic
hydrogen that uses a simple bound-state model for the proton, with only
one free parameter, the radius of the well, which we have chosen to be
$R=1.2$ fm. Once the formalism is set up, standard techniques of quantum
field theory can be used to evaluate proton structure effects using this
parameter, in contrast to the standard approach, which involves the
analysis of forward photon proton scattering. We do not claim our
approach is better, only that it introduces a different way of looking
at the problem. 

We have considered only Coulomb photons in this paper. This is because
our primary concern was in setting up the basic formalism, and testing
it in the relatively tractable case of Coulomb photon exchange. To
continue, while one could stay in the Coulomb gauge and introduce
transverse photons, it is simpler to simply change to the Feynman gauge.
Because the quarks all have the same energy, much of the work here is
effectively already in the Feynman gauge, with
$\gamma_{\mu}\cdots\gamma^{\mu}  \rightarrow \gamma_0\cdots\gamma_0.$
Continuing the calculation to this more complete state will allow the
treatment of the interesting case of hyperfine splitting.  The
difficulties of proton structure are well known to be exacerbated in
this case, with even ground-state hydrogen hfs uncertain at the sixth
digit. Our approach should allow a direct calculation of what is often
referred to as dynamic proton polarizability, as well as allowing a
systematic treatment of other spin dependent effects that sometimes are
difficult to disentangle. These calculations are complicated by the fact
that the loop energy integral can no longer be evaluated with Cauchy's
theorem. Instead one must carry out a Wick rotation to the imaginary
axis and carry out the integral numerically. This rotation requires care
because of the presence of poles and cuts in the complex plane, which
leads to a number of extra terms.

We have shown that in our method a term, while extremely small, that can
be thought of as the Lamb shift of the proton arises. The smallness of
the effect had very much to do with the fact that all three quarks are
taken to be in the $1s$ state, which limits the $\kappa$ values allowed
for the propagator. It is an interesting open question as to how this
effect would change if corrections to the proton wave function involving
non-$s$ states were present, as they presumably are because of gluon
exchange, but that is outside the scope of this paper.

We have shown that the one-photon diagrams together with the part of the
two-photon diagrams where the proton is unchanged give the result
\begin{equation}
E_{\rm fns}({\mbox{non-recoil}}) = \left[-7.161\,22 \,\tilde r^{2 \gamma} 
+ 0.046\,70 \,\tilde r_p^3\right]~\mbox{meV}.
\end{equation}
This was calculated without the large recoil corrections present in
muonic hydrogen: multiplying the appropriate factors of
$(m_\rr/m_{\rmssmu})^3$ and $(m_\rr/m_{\rmssmu})^4$ into the first and
second terms, our result becomes
\begin{equation}
E_{\rm fns} = \left[-5.199\,5 \, \tilde r^{2 \gamma} 
+ 0.030\,5 \, \tilde r_p^3\right]~\mbox{meV}.
\end{equation}

The polarizability of the proton was calculated in a novel fashion.  The
basic result of this paper is that despite the difference in approach, a
similarly small result is found. Specifically, again restoring recoil
corrections by multiplying our result by $(m_\rr/m_{\rmssmu})^4$, we
have found the proton polarizability correction 
\begin{equation}
\Delta E_{\rm pol} = 0.017 \, {\rm meV}.
\end{equation}
While inclusion of transverse photons  may lead to quantitative changes,
we consider it unlikely that a qualitative change will result.

We now compare our results with Ref.~\cite{Pohl}, where the formula
\begin{equation}
\Delta E_{\rm LS} = 206.0336(15) -5.2275(10) r_p^2 + \Delta E_{\rm TPE}
\end{equation}
is given in their Eq.~(32), with the breakdown of $\Delta
E_{\rm TPE}=0.0351(20)$ meV given in Table I along with results from other
calculations that differ by under 2 $\rmmu$eV. 

The quadratic term compares well when a term $-0.0275$ arising from
radiative corrections to finite size, not treated here, is removed. The
breakup of $\Delta E_{\rm TPE}$ involves three terms, a subtraction term
of $-0.002$ meV, an elastic term of $0.023$ meV, and an inelastic term
of $0.014$ meV. The first has no counterpart in our calculation, but if we identify the
elastic term with our Zemach contribution of $0.020$meV and the
inelastic with our polarization of $0.017$ meV we see fairly good
agreement.

Perhaps the most interesting calculation left undone in this framework
is vacuum polarization.  This of course dominates the muonic hydrogen
$2s_{1/2}-2p_{3/2}$ splitting when an electron is in the vacuum
polarization loop, but in our framework we could also put in the
zero-mass quarks confined in the well into the loop. For an isolated
proton, symmetry arguments \cite{oldprl} show that vacuum polarization
vanishes, but once in an atom the arguments no longer hold, and a finite
effect should be present. The standard approach is to introduce pion
loops, in which case the previously mentioned very small ``hadronic
vacuum polarization'' value of 0.011 meV results, but this approach is
quite different. Such calculations are, however, particularly
challenging because of the high degree of divergence present, which
always presents difficulties for bound-state methods. We are
investigating whether techniques that have proved useful in studying
vacuum polarization effects in atoms \cite{Mohr-Soff}, where careful
grouping of angular momentum contributions allows an accurate treatment
of highly-divergent terms, can be extended to this novel case.

\begin{acknowledgments}
The work of J.G. and J.S was supported in part by NSF grant PHY1068065.
We thank S.A. Blundell and K.T. Cheng for useful discussions.

\end{acknowledgments}

\newpage

\begin{figure}[h]
  \begin{center}
    \subfigure[~Two counterterm diagram]
    {\label{2counterterm}\includegraphics[width=7cm]{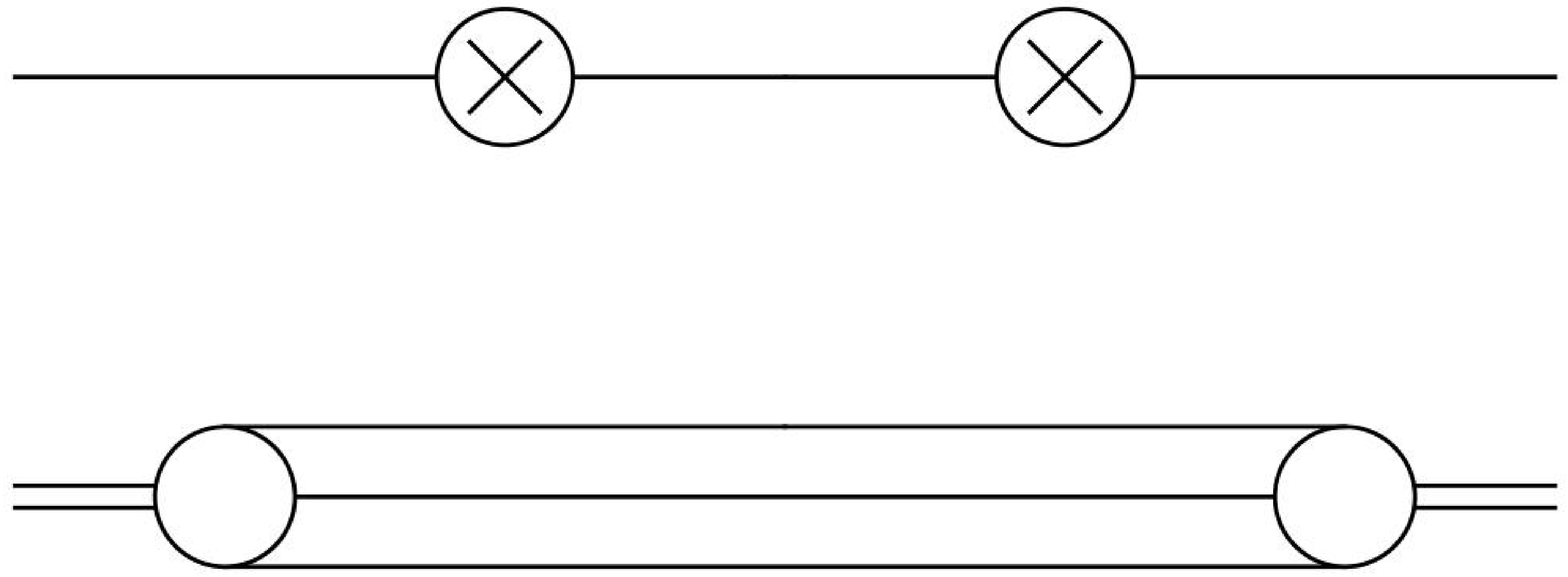}}
    \qquad
    \subfigure[~One counter term one photon exchange diagram]
    {\label{gammacounterterm}\includegraphics[width=7cm]{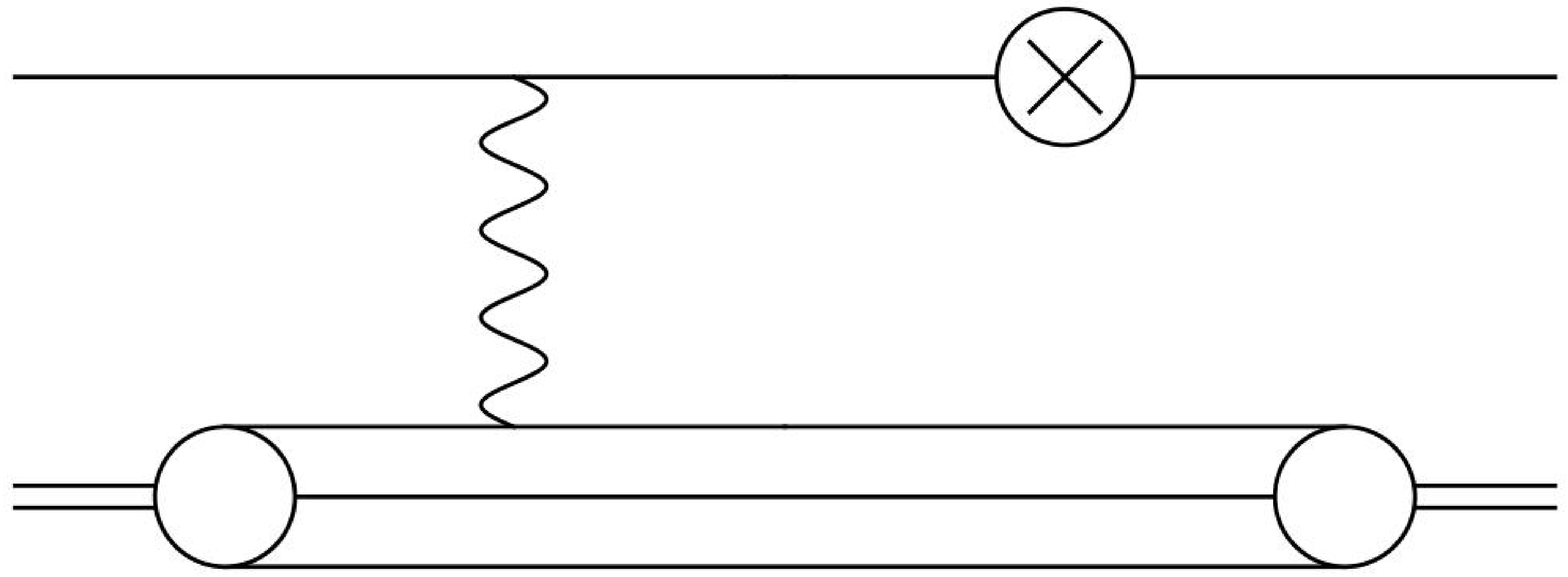}}
    \qquad
      \subfigure[~Zero-loop two photon exchange term]
    {\label{skew}\includegraphics[width=7cm]{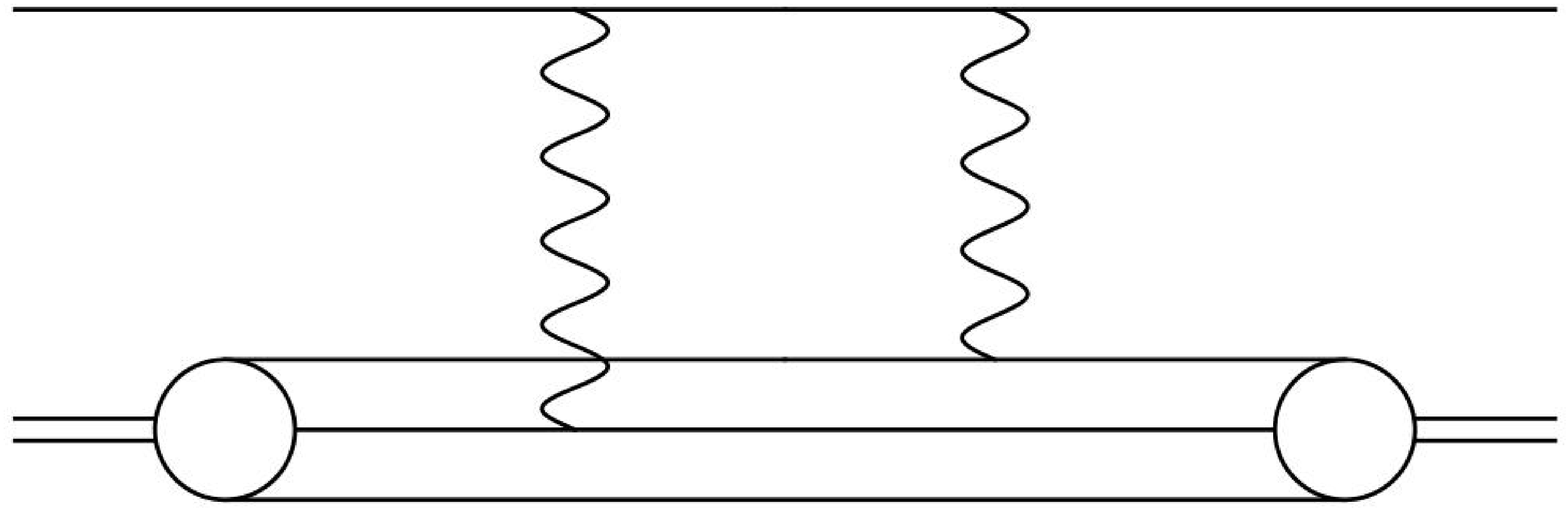}}
\qquad
    \subfigure[~Ladder diagram]
    {\label{ladder}\includegraphics[width=7cm]{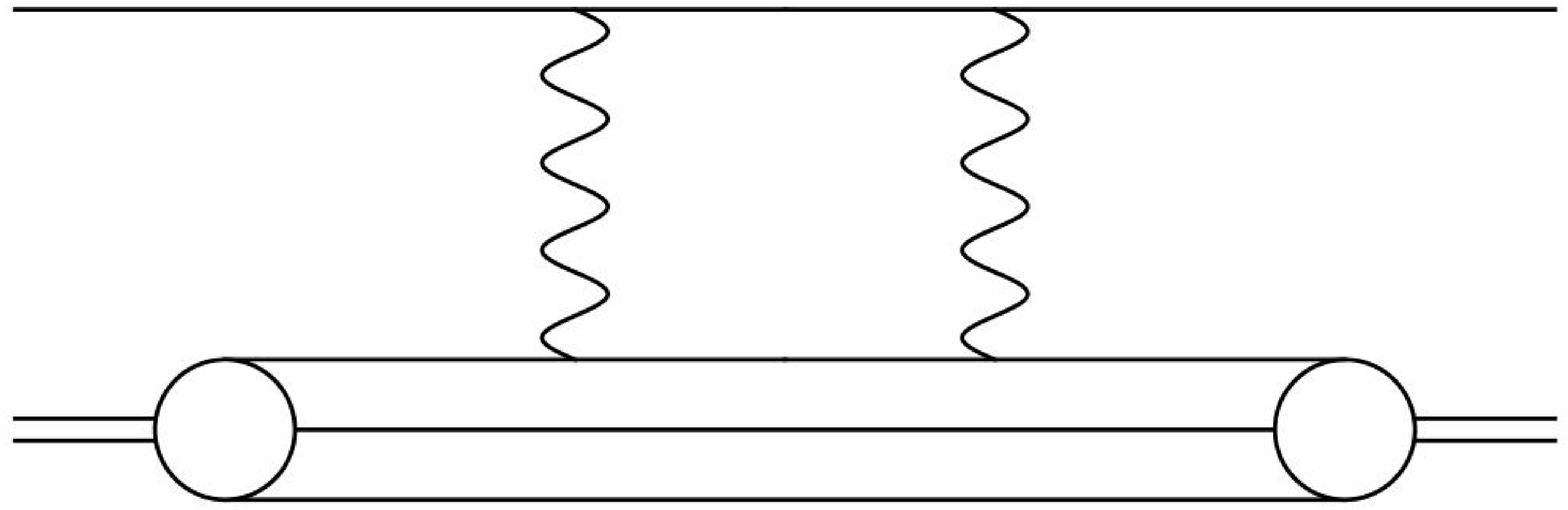}}
    \qquad
    \subfigure[~Crossed ladder diagram]
    {\label{crossedladder}\includegraphics[width=7cm]{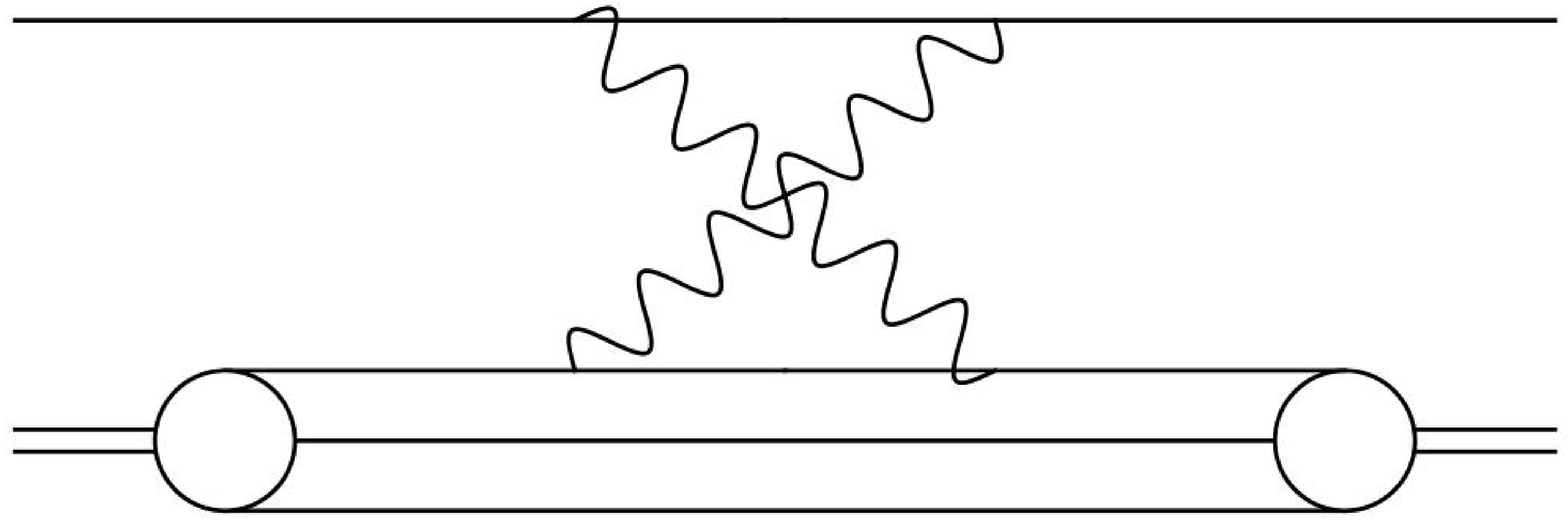}}
   \end{center}
  \caption{Order $\alpha^2$ contributions to polarizability}
  \label{polarization}
\end{figure}
\begin{figure}[h]
  \begin{center}
    \includegraphics[width=7cm]{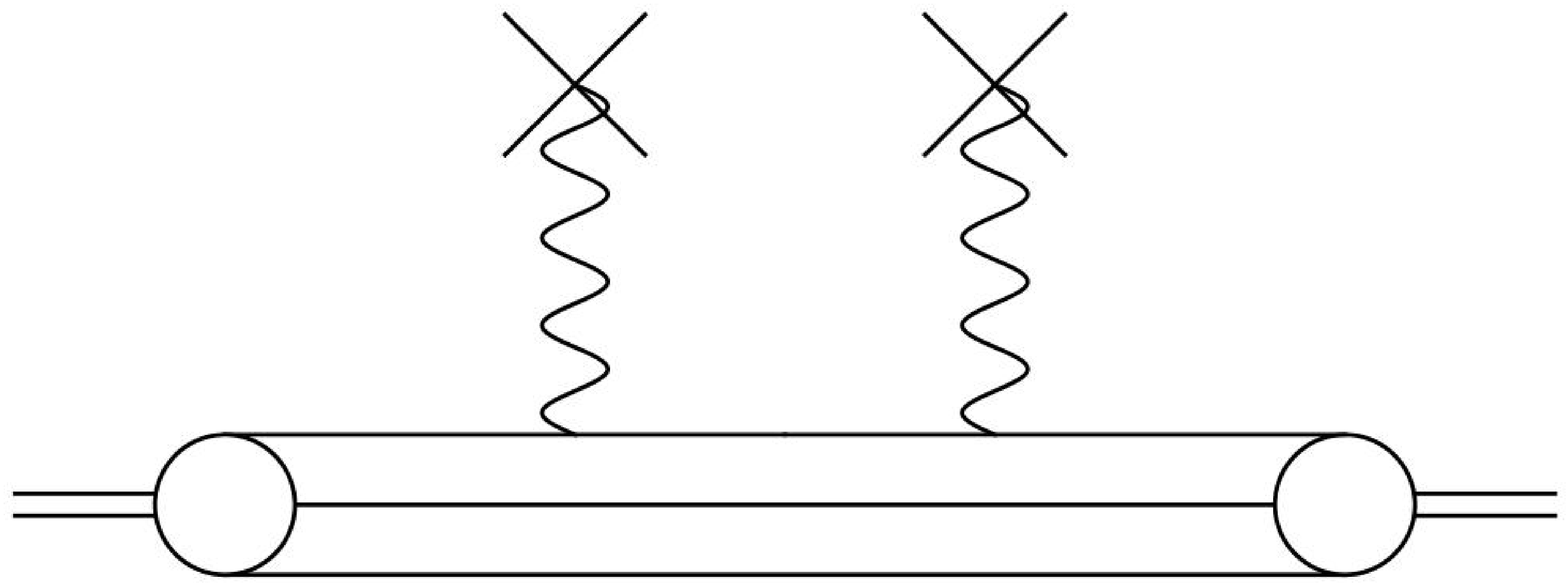}
  \end{center}
  \caption{Diagram for static polarizability}
  \label{Hlamb6}
\end{figure}

\end{document}